\documentclass[ALICE,manyauthors]{cernphprep}

\usepackage[comma,square,numbers,sort&compress]{natbib}
\usepackage{hyperref}
\usepackage{lineno}
\usepackage{float}
\usepackage{comment}
\usepackage{chngcntr}

\begin{document}%

\begin{titlepage}
\PHyear{2018}
\PHnumber{057}      
\PHdate{28 March}  
%

\title{Energy dependence and fluctuations of anisotropic flow\\ in Pb--Pb collisions at $\mathbf{\sqrt{{\textit s}_\text{NN}}} = \mathbf{5.02}$ and $\mathbf{2.76}$ TeV}
\ShortTitle{Fluctuations of elliptic flow}   

\Collaboration{ALICE Collaboration\thanks{See Appendix~\ref{app:collab} for the list of collaboration members}}
\ShortAuthor{ALICE Collaboration} 

\begin{abstract}
Measurements of anisotropic flow coefficients with two- and multi-particle cumulants for inclusive charged particles in Pb--Pb collisions at $\sqrt{{\textit s}_\text{NN}} = 5.02$ and 2.76 TeV are reported in the pseudorapidity range $|\eta| < 0.8$ and transverse momentum $0.2 < p_\text{T} < 50$ GeV/$c$. The full data sample collected by the ALICE detector in 2015 (2010), corresponding to an integrated luminosity of 12.7 (2.0) $\mu$b$^{-1}$ in the centrality range 0--80\%, is analysed. Flow coefficients up to the sixth flow harmonic ($v_6$) are reported and a detailed comparison among results at the two energies is carried out. The $p_\text{T}$ dependence of anisotropic flow coefficients and its evolution with respect to centrality and harmonic number $n$ are investigated. An approximate power-law scaling of the form $v_n(p_\text{T}) \sim p_\text{T}^{n/3}$ is observed for all flow harmonics at low $p_\text{T}$ ($0.2 < p_\text{T} < 3$ GeV/$c$). At the same time, the ratios $v_n/v_m^{n/m}$ are observed to be essentially independent of $p_\text{T}$ for most centralities up to about $p_\text{T} = 10$ GeV/$c$. Analysing the differences among higher-order cumulants of elliptic flow ($v_2$), which have different sensitivities to flow fluctuations, a measurement of the standardised skewness of the event-by-event $v_2$ distribution $P(v_2)$ is reported and constraints on its higher moments are provided. The Elliptic Power distribution is used to parametrise $P(v_2)$, extracting its parameters from fits to cumulants. The measurements are compared to different model predictions in order to discriminate among initial-state models and to constrain the temperature dependence of the shear viscosity to entropy-density ratio. \\ \\ 
\begin{center}
\end{center}
\end{abstract}
\end{titlepage}
\setcounter{page}{2}

\section{Introduction}
The primary goal of ultra-relativistic heavy-ion collisions is to study the properties of QCD matter at extremely high temperatures and/or densities and to understand the microscopic dynamics from which these properties arise, especially in the non-perturbative regime.
The study of anisotropies in the azimuthal distribution of produced particles, commonly called anisotropic flow, has contributed significantly to the characterization of the system created in heavy-ion collisions \cite{Ollitrault:1992bk, Voloshin:2008dg, Alver:2006wh, Alver:2008zza, Alver:2010gr}. According to the current paradigm of bulk particle production, anisotropic flow is determined by the response of the system to its initial spatial anisotropies. Initial-state spatial anisotropies come in turn from both the geometry of the collision and fluctuations in the wave function of the incident nuclei \cite{Alver:2006wh, Alver:2008zza, Alver:2010gr, ALICE:2011ab, ATLAS:2012at, Chatrchyan:2013kba}. The significant magnitude of anisotropic flow is interpreted as evidence of the formation of a strongly-coupled system, which can effectively be described as a fluid with very low shear viscosity to entropy-density ratio ($\eta/s$) \cite{Luzum:2008cw}.

Anisotropic flow is quantified by the coefficients $v_n$ of a Fourier series decomposition of the distribution in azimuthal angle $\varphi$ of final-state particles \cite{Voloshin:1994mz}
\begin{equation}
\frac{\text{d}N}{\text{d}\varphi} \propto 1 + 2 \, \sum_{n=1}^{+\infty} v_n \cos{[n (\varphi - \Psi_n)]},
\end{equation}

where $\Psi_n$ corresponds to the symmetry plane angle of order $n$. The dominant flow coefficient in non-central heavy-ion collisions is the second flow harmonic ($v_2$), called elliptic flow, which is mostly a result of the average ellipsoidal shape of the overlapping area between the colliding nuclei, whereas higher harmonics originate from initial-state fluctuations. For transverse momenta $p_\text{T} \lesssim 3$ GeV/$c$, anisotropic flow is thought to be quantitatively determined by the whole evolution of the system, including the phase of hadronic rescatterings that takes place after chemical freeze-out \cite{Hirano:2005xf}. Flow coefficients have been shown to be sensitive not only to initial-state anisotropies, but also to the transport parameters (such as shear and bulk viscosity \cite{Romatschke:2009im, Shen:2010uy}) and the equation of state of the system, and they have therefore been used to constrain these properties \cite{Ryu:2015vwa, Pratt:2015zsa}. However, given the different heterogeneous phases that the system is believed to undergo, it has not been possible so far to simultaneously constrain the large number of model parameters, although attempts have been made \cite{Bernhard:2016tnd, Auvinen:2017fjw}.

In this regard, the energy dependence of anisotropic flow has been shown to provide additional discriminating power over initial-state models and temperature dependence of transport parameters \cite{Niemi:2015voa, Noronha-Hostler:2015uye}. In fact, some theoretical uncertainties in the determination of anisotropic flow coefficients are expected to partially cancel in the ratios of $v_n$ coefficients measured at different collision energies, such as those on the choice of initial-state model or on the absolute value of $\eta/s$. These ratios would then effectively constrain the variations with collision energy and, therefore, system temperature of the parameters to which anisotropic flow is most sensitive.  

It is known that the magnitude of anisotropic flow, being approximately proportional to the initial-state spatial anisotropy \cite{Gardim:2011xv}, fluctuates from collision to collision even for fixed centrality \cite{Agakishiev:2011eq, Alver:2010rt, Alver:2007qw, Aamodt:2010pa, ALICE:2011ab}, and that its probability distribution function (p.d.f.) $P(v_n)$ is to a first approximation Bessel-Gaussian \cite{Ollitrault:1992bk, Voloshin:2007pc}, i.e.\ the product of a modified Bessel function and a Gaussian function. It has been pointed out that small deviations from a Bessel-Gaussian shape are to be expected independently from the details of initial-state fluctuations \cite{Yan:2013laa, Yan:2014afa, Gronqvist:2016hym}. Evidence of such small deviations has been previously reported \cite{Aad:2014vba}. These deviations are due to first order to the flow p.d.f.\ having a finite skewness. Its quantitative determination would therefore improve the characterization of these deviations. For dimensional reasons, it is convenient to use a standardised skewness ($\gamma_1$), defined as \cite{Giacalone:2016eyu}

\begin{equation} \label{eq:gamma_ori}
\gamma_1 = \frac{\langle ( v_n\{\text{RP}\} - \langle v_n\{\text{RP}\} \rangle )^3 \rangle}{\langle ( v_n\{\text{RP}\} - \langle v_n\{\text{RP}\} \rangle )^2 \rangle^{3/2}},
\end{equation}

where $v_n\{\text{RP}\}$ refers to the anisotropic flow with respect to the reaction plane $\Psi_\text{RP}$, i.e.\ the plane spanned by the impact parameter and the beam axis, and the brackets $\langle \cdots \rangle$ indicate an average over all events. It is worthwhile to note that the symmetry planes $\Psi_n$ do not generally coincide with $\Psi_\text{RP}$ because of initial-state fluctuations.  

A robust experimental method to quantify flow fluctuations is to measure $v_n$ with multi-particle cumulants, which have different sensitivities to the moments of the underlying flow p.d.f. $P(v_n)$

\begin{align}
v_n\{2\} &= \sqrt[2]{\langle v_n^{2} \rangle}, \\
v_n\{4\} &= \sqrt[4]{2 \langle v_n^{2} \rangle^2 - \langle v_n^{4} \rangle}, \\
v_n\{6\} &= \sqrt[6]{\langle v_n^{6} \rangle - 9 \langle v_n^{2} \rangle \langle v_n^{4} \rangle + 12 \langle v_n^{2} \rangle^3}, \\
v_n\{8\} &= \sqrt[8]{\langle v_n^{8} \rangle - 16 \langle v_n^{2} \rangle \langle v_n^{6} \rangle - 18 \langle v_n^{4} \rangle^2 + 144 \langle v_n^{2} \rangle^2 \langle v_n^{4} \rangle - 144 \langle v_n^{2} \rangle^4}.
\end{align}

The number in curly brackets indicates the order of the cumulant.

For elliptic flow, a large difference between $v_2\{2\}$ and $v_2\{4\}$ and approximately equal values of the higher order cumulants ($v_2\{4\}$, $v_2\{6\}$, $v_2\{8\}$) have been previously observed \cite{Abelev:2014mda, Aad:2014vba}, which is indeed consistent with an approximately Bessel-Gaussian flow p.d.f.. However, a fine-splitting of a few percent among the higher order cumulants ($v_2\{4\}$, $v_2\{6\}$, $v_2\{8\}$) has also been reported \cite{Aad:2014vba}, which is thought to be determined by the residual deviations from Bessel-Gaussian shape, in particular a non-zero skewness. A negative value of $\gamma_1$, which corresponds to $P(v_2)$ being left skewed, is expected \cite{Yan:2014afa} from the necessary condition on the initial-state eccentricity $\varepsilon_2 < 1$, which acts as a right cutoff on $P(v_2)$. The Elliptic Power distribution, proposed in  \cite{Yan:2013laa, Yan:2014afa}, was motivated mainly by this observation and it was shown to provide a good description of $P(v_2)$ in a wide centrality range \cite{Yan:2014nsa}. Moreover, $\gamma_1$ has been predicted to increase in absolute value from central to peripheral collisions \cite{Giacalone:2016eyu}, being roughly proportional to $\langle v_2\{\text{RP}\} \rangle$ and being inversely proportional to the square root of the system size \cite{Gronqvist:2016hym}. $\gamma_1$ can be estimated from the fine-splitting among two- and multi-particle cumulants \cite{Giacalone:2016eyu}

\begin{equation} \label{eq:gamma}
\gamma_1^\text{exp} = -6\sqrt{2} v_2\{4\}^{2} \frac{v_2\{4\} - v_2\{6\}}{(v_2\{2\}^2 - v_2\{4\}^2)^{3/2}}.
\end{equation}

It is denoted as $\gamma_1^\text{exp}$ to emphasize that it does not exactly match the definition of $\gamma_1$ given in Eq.\ \ref{eq:gamma_ori}, although the two have been estimated to coincide within a few percents \cite{Giacalone:2016eyu}. The derivation of Eq.\ \ref{eq:gamma} relies on a Taylor expansion of the generating function in powers of the moments, truncated at the order of the skewness. It is experimentally possible to test the validity of this approximation through the universal equality that it implies \cite{Jia:2014pza, Giacalone:2016eyu}

\begin{equation} \label{eq:1}
v_2\{6\} - v_2\{8\} = \frac{1}{11} ( v_2\{4\} - v_2\{6\} ).
\end{equation}

The precision up to which this equality holds depends on the residual contribution of higher central moments of the flow p.d.f., e.g.\ the kurtosis, to the multi-particle cumulants.

At high $p_\text{T}$ ($p_\text{T} \gtrsim 10$ GeV/$c$) the dominant mechanism that determines azimuthal anisotropies of the produced final-state particles is thought to be path-length dependent energy-loss of highly energetic partons \cite{Gyulassy:2000gk, Wang:2000fq, Shuryak:2001me}. Although several experimental observations, such as jet azimuthal anisotropies \cite{Aad:2013sla, Adam:2015mda}, are consistent with this hypothesis, the details of the process are largely unconstrained and measurements of anisotropic flow of high-$p_\text{T}$ particles can help in this regard. Although the mechanism that determines it is fundamentally different, the origin of anisotropic flow at high $p_\text{T}$ is common to the one at low $p_\text{T}$: initial-state geometry and its event-by-event fluctuations. Measurements reported in \cite{Sirunyan:2017pan} seem to confirm this interpretation.

Recent CMS results on non-Gaussian elliptic flow fluctuations \cite{Sirunyan:2017fts} appeared during the writing of this article. Numerical data are not yet available, but the observations seem to be essentially compatible with our measurements and their conclusions agree with those of this article.

\section{Data sample and analysis methods}
The sample of Pb--Pb collisions used for this measurement was recorded with the ALICE detector \cite{Abelev:2014ffa, Aamodt:2008zz} in November and December 2015 (2010), during the Run 2 (Run 1) of the LHC, at a centre of mass energy per nucleon of $\sqrt{{\textit s}_\text{NN}} = 5.02$ (2.76) TeV. The detectors used in the present analysis are the Inner Tracking System (ITS) and Time Projection Chamber (TPC), for primary vertex determination and charged particle tracking, and the V0 detector, for symmetry plane determination, centrality estimation \cite{Abelev:2013qoq} and trigger. The trigger conditions are described in \cite{Abelev:2014ffa}. About $78.4 \times 10^6$ ($12.6 \times 10^6$) minimum-bias events in the centrality range 0--80\%, corresponding to an integrated luminosity of 12.7 $\mu$b$^{-1}$ (2.0 $\mu$b$^{-1}$), with a reconstructed primary vertex position along the beam direction ($z_\text{vtx}$) within $\pm 10$ cm from the nominal interaction point, passed offline selection criteria \cite{Abelev:2014ffa} for the data sample at $\sqrt{{\textit s}_\text{NN}} = 5.02$ (2.76) TeV.
Centrality is determined from the measured amplitude in the V0, which is proportional to the number of charged tracks in the corresponding acceptance ($2.8 < \eta< 5.1$ for V0A and $-3.6 < \eta < -1.7$ for V0C).

Charged tracks with transverse momentum $0.2 < p_\text{T} < 50$ GeV/$c$ and pseudorapidity $|\eta|< 0.8$ are used in the present analysis. These tracks are reconstructed using combined information from the ITS and TPC. A minimum number of TPC space points of 70 (out of 159) is required for all tracks, together with a $\chi^2$ per TPC space point ($\chi^2_\text{TPC}$) in the range $0.1 < \chi^2_\text{TPC} < 4$. A minimum number of 2 ITS hits, of which at least one in the two innermost layers, is required, together with a $\chi^2$ per ITS hit per degree of freedom ($\chi^2_\text{ITS}$) smaller than 36. Only tracks with a distance of closest approach (DCA) to the primary vertex position less than 3.2 cm in the beam direction and 2.4 cm transverse to it are used. These track selection criteria ensure an optimum rejection of secondary particles and a $p_\text{T}$ resolution better than 5\% in the $p_\text{T}$ range used in the present analysis \cite{Abelev:2014ffa}.

Anisotropic flow coefficients are measured with the $Q$-cumulant method \cite{Bilandzic:2010jr}, using the implementation proposed in \cite{Bilandzic:2013kga}. Track weights ($w$) are used in the construction of the $Q$-vectors, in order to correct for non-uniform reconstruction efficiency and acceptance
\begin{equation} \label{eq:Qvec}
 Q_{n,m} = \sum_{j=1}^{M} w_j (p_\text{T}, \eta, \varphi, z_\text{vtx})^m e^{i n \varphi_j},
\end{equation}

where $M$ is the charged track multiplicity, $n$ the harmonic and $m$ an integer exponent of the weights. After applying track weights, the effects due to non-uniformities in azimuthal acceptance, which would introduce a bias in the measured flow coefficients, are observed to be negligible. This is evaluated by measuring the event-averaged values of the real and imaginary part of $Q_{n}$, which are consistent with zero. Multi-particle cumulants are measured on an event-by-event basis and then, in order to minimise statistical fluctuations, averaged over all events using the corrected charged track multiplicity as a weight, following the procedure proposed in \cite{Bilandzic:2010jr}. All observables are computed in small centrality bins (1\%) and then integrated, when limited size of the data sample makes it necessary, in wider centrality intervals using the charged particle yield in each 1\% centrality bin as weight. This avoids that the event weighting procedure, based on multiplicity, would introduce a bias in the average centrality within a large centrality bin, since multiplicity varies with centrality.

For $p_\text{T}$-integrated results, the $m$-particle cumulants are calculated using all tracks within given $p_\text{T}$ range, while for $p_\text{T}$-differential results one particle at a given $p_\text{T}$ is correlated with $m-1$ particles in the full $p_\text{T}$ range ($0.2 < p_\text{T} < 50$ GeV/$c$). In terms of reference ($c_n\{m\}$) and differential ($d_n\{m\}$) cumulants, as defined in \cite{Bilandzic:2010jr}, the flow coefficients are measured as

\begin{align}
&v_n\{2\} = \sqrt[2]{c_n\{2\}}, \\
&v_n\{4\} = \sqrt[4]{- c_n\{4\}}, \\
&v_n\{6\} = \sqrt[6]{\frac{1}{4} c_n\{6\}}, \\
&v_n\{8\} = \sqrt[8]{-\frac{1}{33} c_n\{8\}}, \\
&v_n\{2\}(p_\text{T}) = d_n\{2\}(p_\text{T}) / \sqrt[2]{c_n\{2\}}, \\
&v_n\{4\}(p_\text{T}) = - d_n\{4\}(p_\text{T}) / \sqrt[4]{- c_n\{4\}^3 }.
\end{align}

For two-particle correlations, a separation in pseudorapidity between the correlated particles ($\Delta\eta$) is applied in order to suppress short-range azimuthal correlations which are not associated to the symmetry planes, usually called `non-flow'. These correlations arise from jets, mini-jets and resonance decays. For flow coefficients of higher order ($v_n\{m>2\}$), non-flow contribution has been previously found to be negligible in Pb--Pb collisions \cite{Aamodt:2010pa, Abelev:2014mda}. Results corresponding to $|\Delta\eta| > 1$ (denoted with $v_n\{2, |\Delta\eta|>1\}$) are obtained with the two-particle cumulant correlating tracks from opposite sides of the TPC acceptance, $-0.8 < \eta < -0.5$ and $0.5 < \eta < 0.8$.
Results corresponding to $|\Delta\eta| > 2$ (and reported as $v_n\{2, |\Delta\eta|>2\}$) are obtained with the scalar product method \cite{Adler:2002pu}, correlating all tracks at mid-rapidity ($|\eta|< 0.8$) with the $n$-th harmonic $Q$-vector $Q_n^\text{V0A}$ calculated from the azimuthal distribution of the energy deposition measured in the V0A detector \cite{Voloshin:2008dg, Luzum:2012da}
\begin{equation}
v_n\{2, |\Delta\eta|>2\} = \frac{\langle u_{n,0} Q_n^\text{V0A*} \rangle}{\sqrt{\frac{\langle Q_n^\text{V0A} Q_{n,1}^*\rangle \langle Q_n^\text{V0A} Q_n^\text{V0C*}\rangle}{\langle Q_{n,1} Q_n^\text{V0C*} \rangle}}},
\end{equation}
where $u_{n,0} = e^{in\varphi}$ is the unit flow vector from charged particle tracks at mid-rapidity and $Q_{n,1}$ is computed from the same type of tracks according to Eq.\ \ref{eq:Qvec}. Both methods have their own limitations and thus are complementary to each other: $v_n\{2, |\Delta\eta|>2\}$ can be reliably employed only up to the fourth harmonic, because of the finite azimuthal segmentation of the V0 detectors (8 sectors in $2\pi$), while $v_n\{2, |\Delta\eta|>1\}$ suffers from bigger statistical uncertainties, due to the limited acceptance from which tracks are selected, and bigger non-flow contribution for $p_\text{T}>10$ GeV/$c$.


The systematic uncertainties are evaluated by varying the track and event selection criteria and comparing the variation in the flow coefficients relative to the default results. The absolute value of the variation itself is assigned as a systematic uncertainty if it is considered significant according to the Barlow criterion \cite{Barlow:2002yb}. Different track quality variables are varied: number of TPC space points, $\chi^2_\text{TPC}$ and $\chi^2_\text{ITS}$, fraction of shared TPC space points and number of ITS hits. For each of these, the default values are varied in order to increase the fraction of excluded tracks as much as 5 times. No significant differences are observed in the reported measurements between positively and negatively charged particles. Concerning the event selection criteria, the following are investigated: polarity of the magnetic field, reconstructed primary vertex position along the beam direction (selecting only events with z$_\text{vtx}$ within $\pm 8$ cm from the nominal interaction point), pile-up rejection (imposing stronger or weaker constraints on the consistency of different event multiplicity estimators) and variations in the instantaneous luminosity delivered to the ALICE detector by the LHC. The uncertainty on centrality determination is evaluated using an alternative estimator based on the number of hits in the second ITS layer ($|\eta| < 1.4$), which is directly proportional to the number of charged particles in the corresponding acceptance. Among the aformentioned sources, for all observables in this article, track quality and centrality determination are the dominant sources. The total systematic uncertainties are evaluated summing in quadrature the systematic uncertainties coming from each of the sources, i.e.\ considering the different sources to be uncorrelated.

\section{Collision energy, transverse momentum and centrality dependence}

\begin{figure}
\centering
  \includegraphics[width=0.6\linewidth]{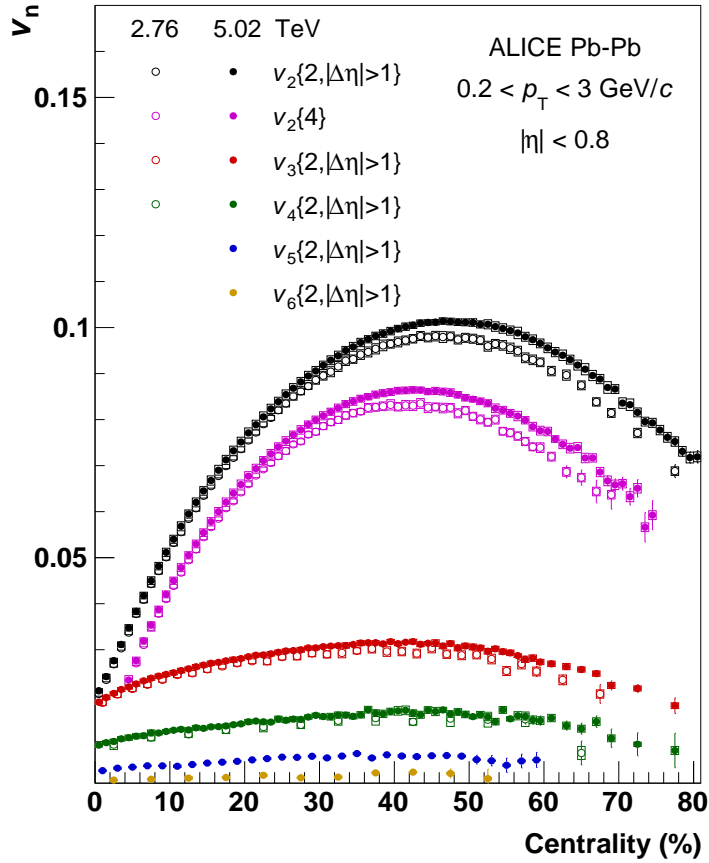}
  \caption{Anisotropic flow coefficients $v_n$ of inclusive charged particles as a function of centrality, for the two-particle (denoted with $|\Delta\eta|>1$) and four-particle cumulant methods. Measurements for Pb--Pb collisions at $\sqrt{{\textit s}_\text{NN}} = 5.02$ (2.76) TeV are shown by solid (open) markers.}
  \label{fig:1}
\end{figure}

Figure \ref{fig:1} presents the centrality dependence of the flow coefficients $v_{n}$  $(n=2, \dots 6)$ averaged in the $p_\text{T}$ range $0.2 < p_\text{T} < 3.0$ GeV/$c$, where collective effects are expected to dominate azimuthal anisotropies. Measurements are performed with the two- and four-particle cumulant method, denoted with $v_n\{2,|\Delta\eta|>1\}$ and $v_n\{4\}$, respectively. Results at both $\sqrt{{\textit s}_\text{NN}} = 5.02$ and $2.76$ TeV are shown. A clear hierarchy is observed among flow coefficients, with the second harmonic (elliptic flow) being the dominant one and the higher harmonics progressively smaller. The centrality-averaged (0--50\%) values of harmonics $v_3-v_6$ are decreasing as $v_{n+1}/v_{n} \sim 0.5$. In contrast to a strong increase in $v_2$ from central to mid-central collisions and a decrease after about $45\%$ centrality towards peripheral collisions, a weak centrality dependence is observed for the higher harmonics. This holds true at both energies and is consistent with previous observations \cite{Aamodt:2010pa, Adam:2016izf}. The characteristic centrality dependence of the elliptic flow was observed already at RHIC energies \cite{Ackermann:2000tr}. Compared to previous measurements in the $p_\text{T}$ range $0.2 < p_\text{T} < 5$ GeV/$c$ \cite{Adam:2016izf}, the differences in $v_n$ coefficients arising from the different choice of $p_\text{T}$ range are of about 1\% and 2\% for $v_2$ and $v_3-v_6$, respectively. 

\begin{figure}
  \centering
  \includegraphics[width=0.6\linewidth]{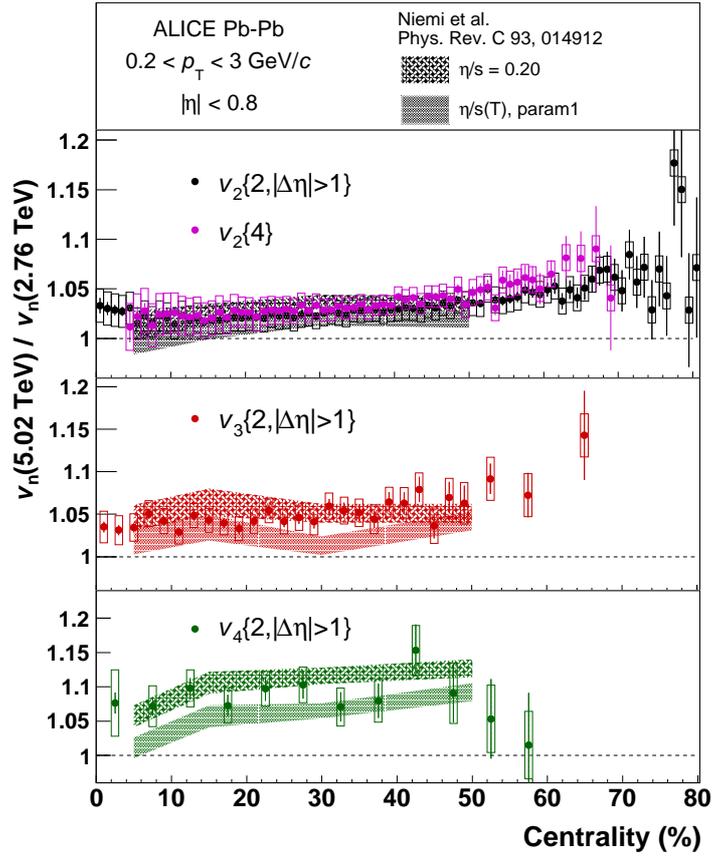}
  \caption{Ratios of anisotropic flow coefficients $v_n$ of inclusive charged particles between Pb--Pb collisions at $\sqrt{{\textit s}_\text{NN}} = 5.02$  and 2.76 TeV, as a function of centrality. Hydrodynamic calculations employing different $\eta/s(T)$ parametrizations \cite{Niemi:2015voa} are shown for comparison.}
  \label{fig:2}
\end{figure}

Figure \ref{fig:2} shows the ratio of $v_{n}\{2,|\Delta\eta|>1\}$ $(n=2,3,4)$ and $v_2\{4\}$ between $\sqrt{{\textit s}_\text{NN}} = 5.02$ and $2.76$ TeV, i.e.\ the relative variation of these flow coefficients between those two energies. 
Since the systematic uncertainties of the measurements at different energies are partially correlated, the resulting systematic uncertainty on the ratio is reduced.
All harmonics are observed to increase with energy, between about 2 and 10\%. A hint of a centrality dependence is observed only for $v_2$, with the increase growing slightly from mid-central towards more peripheral collisions. No significant difference is observed in the increase of $v_2$ measured with two- or four-particle correlations. Since the difference between $v_2\{2,|\Delta\eta|>1\}$ and $v_2\{4\}$ is directly related to flow fluctuations, this observation suggests that the fluctuations of elliptic flow do not vary significantly between the two energies, within experimental uncertainties. The ratios are compared to hydrodynamical calculations with EKRT initial conditions \cite{Eskola:1999fc} and different parametrisations of the temperature dependence of $\eta/s$ \cite{Niemi:2015voa}. The $p$-values for the comparison between data and models are also shown in Tab.\ \ref{tab:chi2}. Among the two parametrisations that provide the best description of RHIC and LHC data \cite{Niemi:2015qia}, both are consistent with the measurements, except for $v_3\{2,|\Delta\eta|>1\}$, albeit the one with constant $\eta/s = 0.2$ agrees slightly better. These comparisons take into account the correlation between systematic uncertainties of data points in different centrality intervals. This observation might indicate little or no temperature dependence of $\eta/s$ within the temperature range at which anisotropic flow develops at the two center of mass energies. As a reference, the $p$-values for the comparison between data and unity in the same centrality range (5--50\%) are also reported in Tab.\ \ref{tab:chi2}. 

\begin{table}
\centering
\begin{tabular}{c|c|c|c}
                          & $\eta/s=0.2$ & $\eta/s(T)$ param1 & 1 \\ \hline
$v_{2}\{2,|\Delta\eta|>1\}$ &    0.712   &  0.645  &  0.477            \\ \hline
$v_{2}\{4\}$                &    0.467   &  0.357  &  0.028           \\ \hline
$v_{3}\{2,|\Delta\eta|>1\}$ &    0.053   &  0.003  &  0.001           \\ \hline
$v_{4}\{2,|\Delta\eta|>1\}$ &    0.484   &  0.468  &  0.022          
\end{tabular}
\caption{$p$-values for the comparison among ratios of $v_{n}\{2,|\Delta\eta|>1\}$ $(n=2,3,4)$ and $v_2\{4\}$ between $\sqrt{{\textit s}_\text{NN}} = 5.02$ and $2.76$ TeV and model calculations using different parametrisations of $\eta/s(T)$ \cite{Niemi:2015voa}, shown in Fig.\ \ref{fig:2}, and unity, in the centrality range 5-50\%.}
\label{tab:chi2}
\end{table}

\begin{figure}
\centering
  \includegraphics[width=\linewidth]{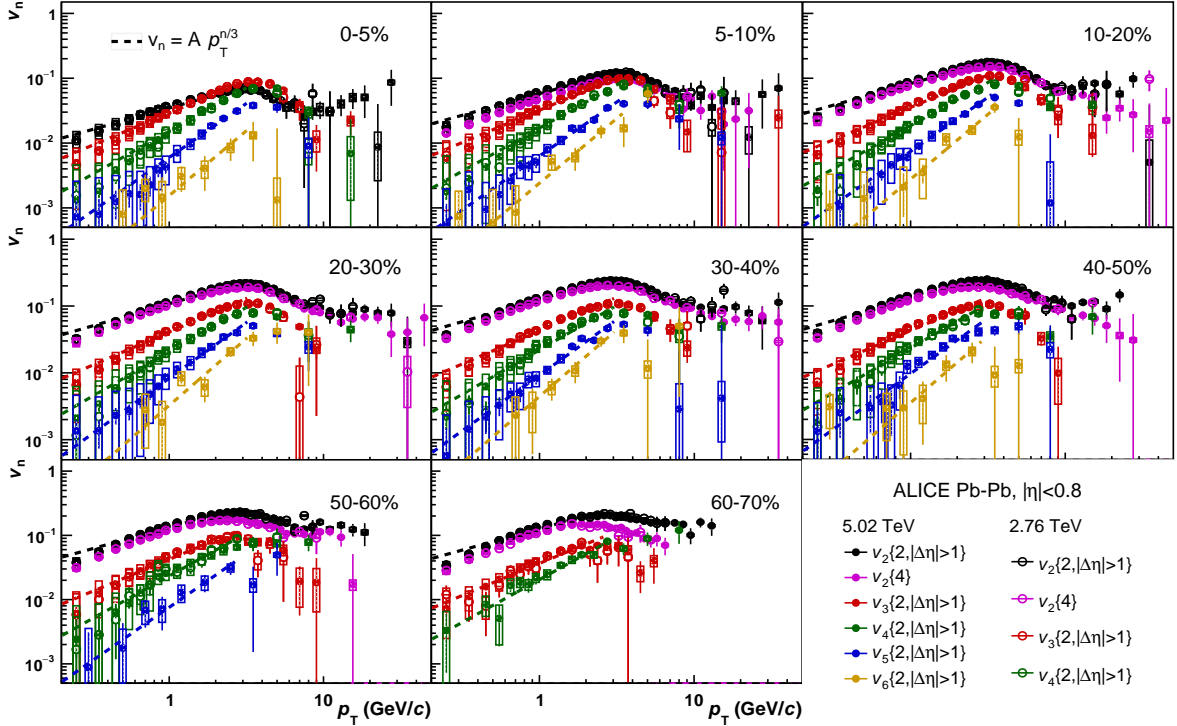}
  \caption{Anisotropic flow coefficients $v_n(p_\text{T})$ of inclusive charged particles in different centrality classes, measured with two-particle (denoted with $|\Delta\eta|>1$) and four-particle cumulant methods. Measurements for Pb--Pb collisions at $\sqrt{{\textit s}_\text{NN}} = 5.02$ (2.76) TeV are shown by solid (open) markers. Dashed lines are fits with a power-law function $v_n(p_\text{T}) = \text{A} \, p_\text{T}^{n/3}$, with A as free parameter.}
  \label{fig:7}
\end{figure}

Figure \ref{fig:7} shows the $p_\text{T}$-differential measurements of $v_n$ ($n = 2, \dots 6$), with two- and four-particle cumulants, in the $p_\text{T}$ range $0.2 < p_\text{T} < 50$ GeV/$c$ and in wide centrality bins, between 0\% and 70\%. The $p_\text{T}$ dependence is qualitatively similar for all harmonics: $v_n$ increases with increasing $p_\text{T}$ up to about 3--4 GeV/$c$, after which it starts decreasing. Comparing the different harmonics, they seem to follow the hierarchy observed in the $p_\text{T}$-integrated results in the whole $p_\text{T}$ range: $v_2 > v_3 > \dots v_6$, except for very central collisions (0--5\%), where $v_3\{2,|\Delta\eta|>1\}$ is observed to be greater than $v_2\{2,|\Delta\eta|>1\}$ for $p_\text{T} \gtrsim 2$ GeV/$c$ and $v_4\{2,|\Delta\eta|>1\}$ is observed to be similar to $v_2\{2,|\Delta\eta|>1\}$ for $p_\text{T} \gtrsim 3$ GeV/$c$. In the centrality range 10--40\% a significant non-zero value of $v_2\{2,|\Delta\eta|>1\}$ and $v_2\{4\}$ is observed up to $p_\text{T} \approx 30$ GeV/$c$; for the higher harmonics, significant values are only measured for $p_\text{T} \leq 10$ GeV/$c$.

Looking at the $p_\text{T}$ dependence in more detail, the flow harmonics are found to follow an approximate power-law scaling up to around the maximum, with exponents being proportional to the harmonic number $n$, $v_n(p_\text{T}) \sim p_\text{T}^{n/3}$, as shown by the dashed fitted lines in Fig.\ \ref{fig:7}. In ideal hydrodynamics, the $p_\text{T}$ dependence of anisotropic flow for massive particles should follow a power-law function $v_n(p_\text{T}) \sim p_\text{T}^n$ in the region of $p_\text{T}/M$ up to order one, where $M$ is the particle's mass, and at higher momenta it has been predicted to be linear in $p_\text{T}$ for all n, $v_n(p_\text{T}) \sim p_\text{T}$ \cite{Borghini:2005kd, Alver:2010dn}. This $p_\text{T}$ dependence is notably different from the one observed in the data. At very low $p_\text{T}$ this is presumably because the relevant momentum region for inclusive particles, mostly pions, is below the range of our measurements, and at higher $p_\text{T}$ ideal hydro is not expected to hold because of momentum dependent viscous corrections at freeze out \cite{Teaney:2003kp} and/or non-linear mode mixing for $n \geqslant 4$ \cite{Qiu:2011iv, Gardim:2011xv}. The power-law dependence for $n=2$ was noticed before and it was attributed to a novel energy loss mechanism \cite{Andres:2016mla}, which however cannot explain the scaling observed for $n>2$. The emergence of this simple power-law dependence remains unexpected and surprising. 

\begin{figure}
\centering
  \includegraphics[width=\linewidth]{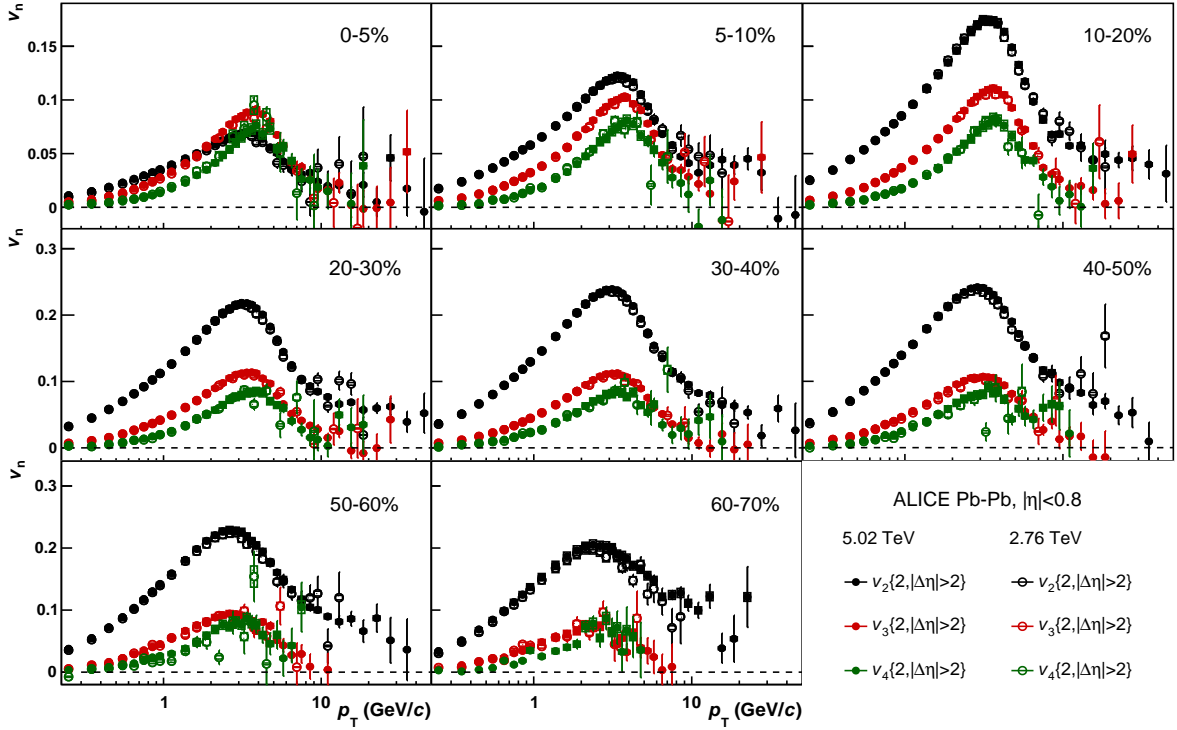}
  \caption{Anisotropic flow coefficients $v_n(p_\text{T})$ of inclusive charged particles in different centrality classes, measured with the scalar product method with respect to the V0A $Q$-vector. Measurements for Pb--Pb collisions at $\sqrt{{\textit s}_\text{NN}} = 5.02$ (2.76) TeV are shown by solid (open) markers.}
  \label{fig:8}
\end{figure}

Figure \ref{fig:8} shows the $p_\text{T}$-differential measurements of $v_n$ ($n=2,3,4$) calculated with the scalar product method with respect to V0A. The same $p_\text{T}$ and centrality range as in Fig.\ \ref{fig:7} is shown. A significant $v_2\{2,|\Delta\eta|>2\}$ is observed up to $p_\text{T} \approx 40$ GeV/$c$ in the centrality range 10--40\%. $v_{n}\{2,|\Delta\eta|>2\}$ and $v_{n}\{2,|\Delta\eta|>1\}$ $(n=2,3,4)$ are found to be compatible within 2\% in the $p_\text{T}$ range $0.2 < p_\text{T} < 10$ GeV/$c$, while a systematic difference (with $v_2 \{2, |\Delta \eta|>1 \} > v_2 \{2, |\Delta \eta|>2 \}$) is observed for $10 < p_\text{T} < 50$ GeV/$c$, ranging from about 3\% in centrality 0--5\% to about 10\% in centrality 40--50\%. This difference is attributed to small residual non-flow contributions which are suppressed by the larger pseudorapidity gap. Two-particle non-flow contributions roughly scale as the inverse of the multiplicity \cite{Voloshin:2008dg}, which is consistent with the observed centrality dependence. Possible differences among $v_{n}\{2,|\Delta\eta|>1\}$ and $v_{n}\{2,|\Delta\eta|>2\}$ $(n=2,3,4)$ arising from the decorrelation of event planes at different pseudorapidities have been estimated to be less than 1\% and 3\% for $v_2$ and $v_{3-4}$, respectively, based on $\eta$-dependent factorization ratios \cite{Gardim:2012im} measured at 2.76 TeV \cite{Khachatryan:2015oea}. This estimation assumes that such decorrelation only depends on $|\Delta\eta|$ and not $\eta$ in the pseudorapidity range under consideration ($|\eta| < 5.1$).

\begin{figure}
\centering
  \includegraphics[width=0.8\linewidth]{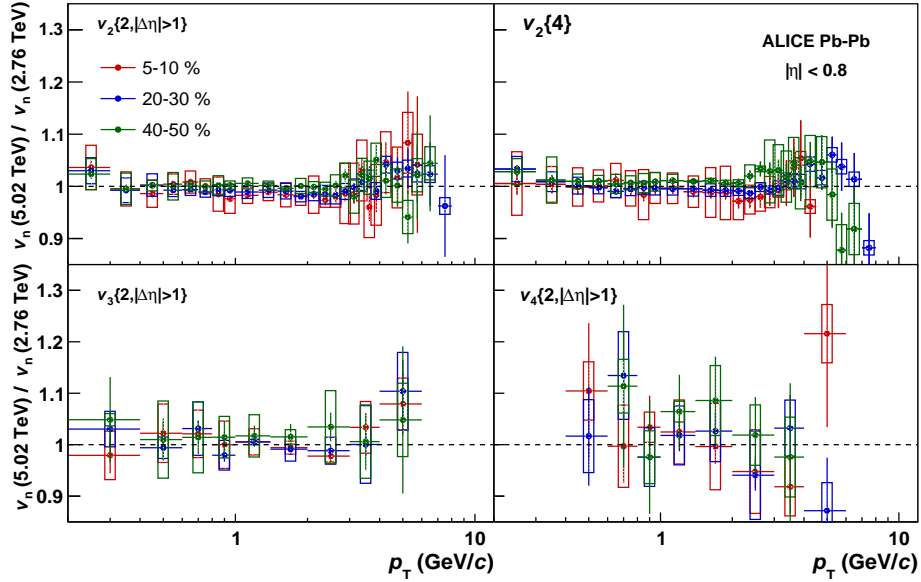}
  \caption{Ratios of anisotropic flow coefficients $v_n(p_\text{T})$ of inclusive charged particles between Pb--Pb collisions at $\sqrt{{\textit s}_\text{NN}} = 5.02$  and 2.76 TeV, in different centrality classes, measured with two-particle (denoted with $|\Delta\eta|>1$) and four-particle cumulant methods.}
  \label{fig:9}
\end{figure}

Figure \ref{fig:9} shows the ratios of $p_\text{T}$-differential $v_{n}\{2,|\Delta\eta|>1\}$  $(n=2,3,4)$ and $v_2\{4\}$ between $\sqrt{{\textit s}_\text{NN}} = 5.02$ and $2.76$ TeV. Overall, the ratios are consistent with unity, indicating that $p_\text{T}$-differential anisotropic flow does not change significantly across collision energies and that the increase observed in the $p_\text{T}$-integrated values can be mostly attributed to an increase of $\langle p_\text{T} \rangle$, as previously noted \cite{Adam:2016izf}. This observation is also consistent with little or no variation of $\eta/s$ between the two collision energies, as already shown in Fig.\ \ref{fig:2}. The possible variations in $p_\text{T}$-integrated values arising from the differences in the $p_\text{T}$-differential ones have been estimated to be less than 1\%, by integrating $v_n(p_\text{T})$ at $\sqrt{{\textit s}_\text{NN}} = 5.02$ ($2.76$) TeV with charged particle spectra at 2.76 (5.02) TeV.

\begin{figure}
\centering
  \includegraphics[width=0.8\linewidth]{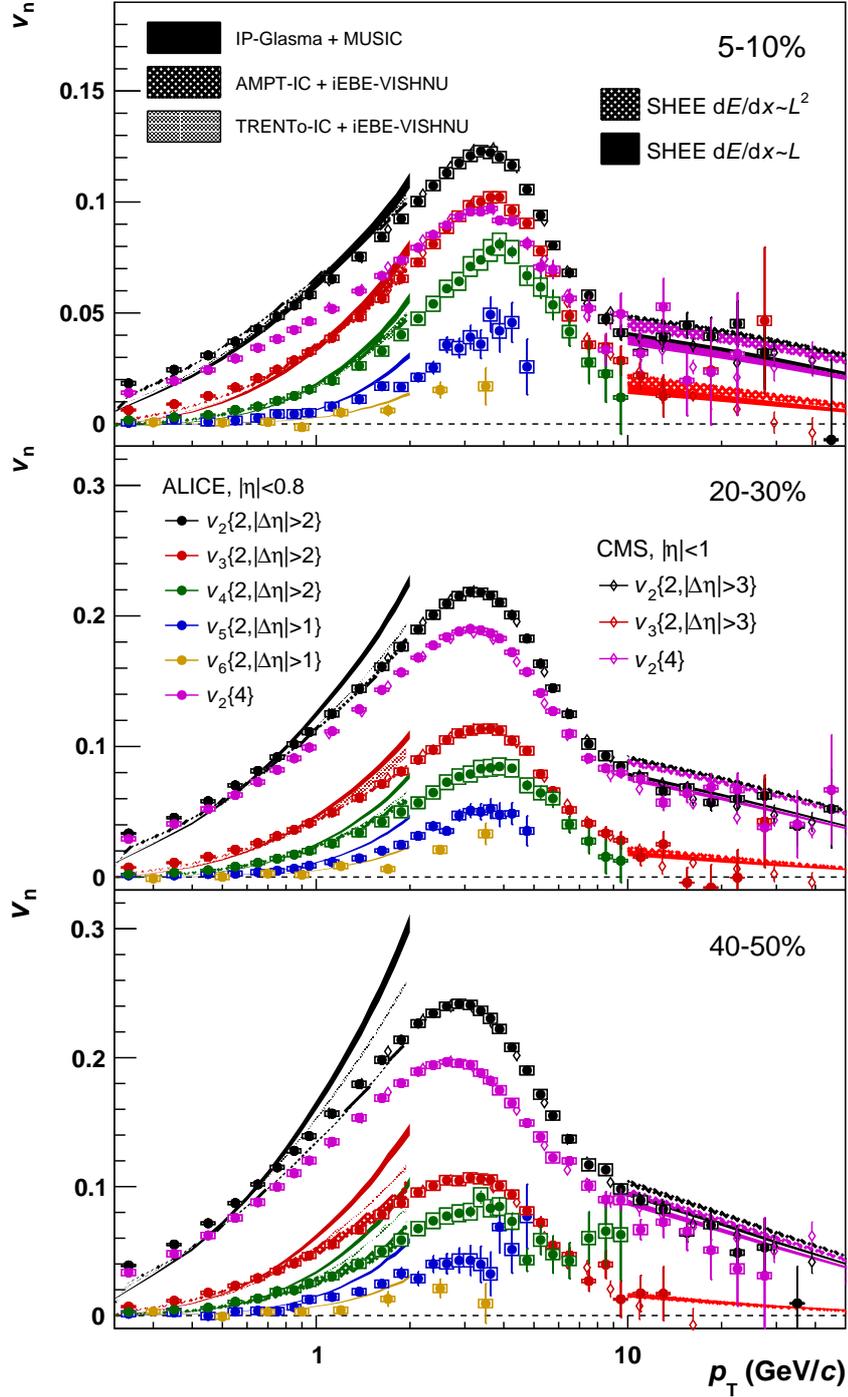}
  \caption{Anisotropic flow coefficients $v_n(p_\text{T})$ of inclusive charged particles in different centrality classes, measured with two- and four-particle cumulant and scalar product methods with respect to the V0A $Q$-vector, for Pb--Pb collisions at $\sqrt{{\textit s}_\text{NN}} = 5.02$ TeV. Several hydrodynamic calculations \cite{McDonald:2016vlt, Zhao:2017yhj, Betz:2016ayq} and previous measurements from CMS \cite{Sirunyan:2017pan} are shown for comparison.}
  \label{fig:hydro}
\end{figure}

Figure \ref{fig:hydro} shows the comparison of $p_\text{T}$-differential flow measurements with different models, in three centrality intervals: 5--10\% (top panel), 20--30\% (middle panel) and 40--50\% (bottom panel). At low $p_\text{T}$ ($p_\text{T} < 2$ GeV/$c$), flow coefficients are expected to be mostly determined by the collective expansion of the system, which is commonly described by hydrodynamic models. The measurements are compared to three calculations, one employing IP-Glasma initial conditions \cite{Bartels:2002cj} matched to the MUSIC viscous hydrodynamic code \cite{McDonald:2016vlt} and two calculations using iEBE-VISHNU viscous hydrodynamic code \cite{Zhao:2017yhj} with AMPT \cite{Lin:2004en} or TRENTo \cite{Moreland:2014oya} initial conditions. The parameters of TRENTo were tuned to reproduce previous measurements in Pb--Pb collisions at $\sqrt{{\textit s}_\text{NN}} = 2.76$ TeV \cite{Bernhard:2016tnd}; with such tuning TRENTo has been shown \cite{Moreland:2014oya} to effectively mimic IP-Glasma initial conditions and therefore the two calculations TRENTo+iEBE-VISHNU and IP-Glasma+MUSIC are expected to be based on similar initial conditions. All models employ a transport cascade model (UrQMD \cite{Bass:1998ca}) to describe the hadronic phase after freeze-out. Compared to data, all models are found to  underestimate the data for $p_\text{T} < 0.5$ GeV/$c$. For $1 < p_\text{T} < 2$ GeV/$c$ the predictions from IP-Glasma+MUSIC and TRENTo+iEBE-VISHNU overestimate the data, while those from AMPT-IC+iEBE-VISHNU are found to be still in agreement. Overall, all models qualitatively describe the $p_\text{T}$ dependence of flow coefficients in this low $p_\text{T}$ range.

At high $p_\text{T}$ ($p_\text{T} > 10$ GeV/$c$), azimuthal anisotropies are on the contrary expected to be determined by path-length dependent parton energy-loss. The measurements are compared to predictions from \cite{Betz:2016ayq}, which combine an event-by-event hydrodynamic description of the medium (v-USPhydro \cite{Noronha-Hostler:2013gga}) with a jet quenching model (BBMG \cite{Betz:2011tu}). Two sets of predictions for $v_{2}\{2\}$, $v_{2}\{4\}$ and $v_{3}\{2\}$, assuming a linear (d$E$/d$x \sim L$) and a quadratic (d$E$/d$x \sim L^2$) dependence of the energy loss on the path length $L$, are compared to data. Other parameters of the model, such as $\eta/s$, are expected to have a minor contribution within the presented centrality ranges \cite{Betz:2016ayq}. For $v_{2}\{2,|\Delta\eta|>2\}$, the linear case is compatible with the data, while the quadratic one can be excluded within 95\% confidence level. For $v_{3}\{2,|\Delta\eta|>2\}$, neither of the two sets of predictions can be excluded within uncertainties. Our results are found to be in good agreement with CMS data \cite{Sirunyan:2017pan}.

\begin{figure}
\centering
  \includegraphics[width=\linewidth]{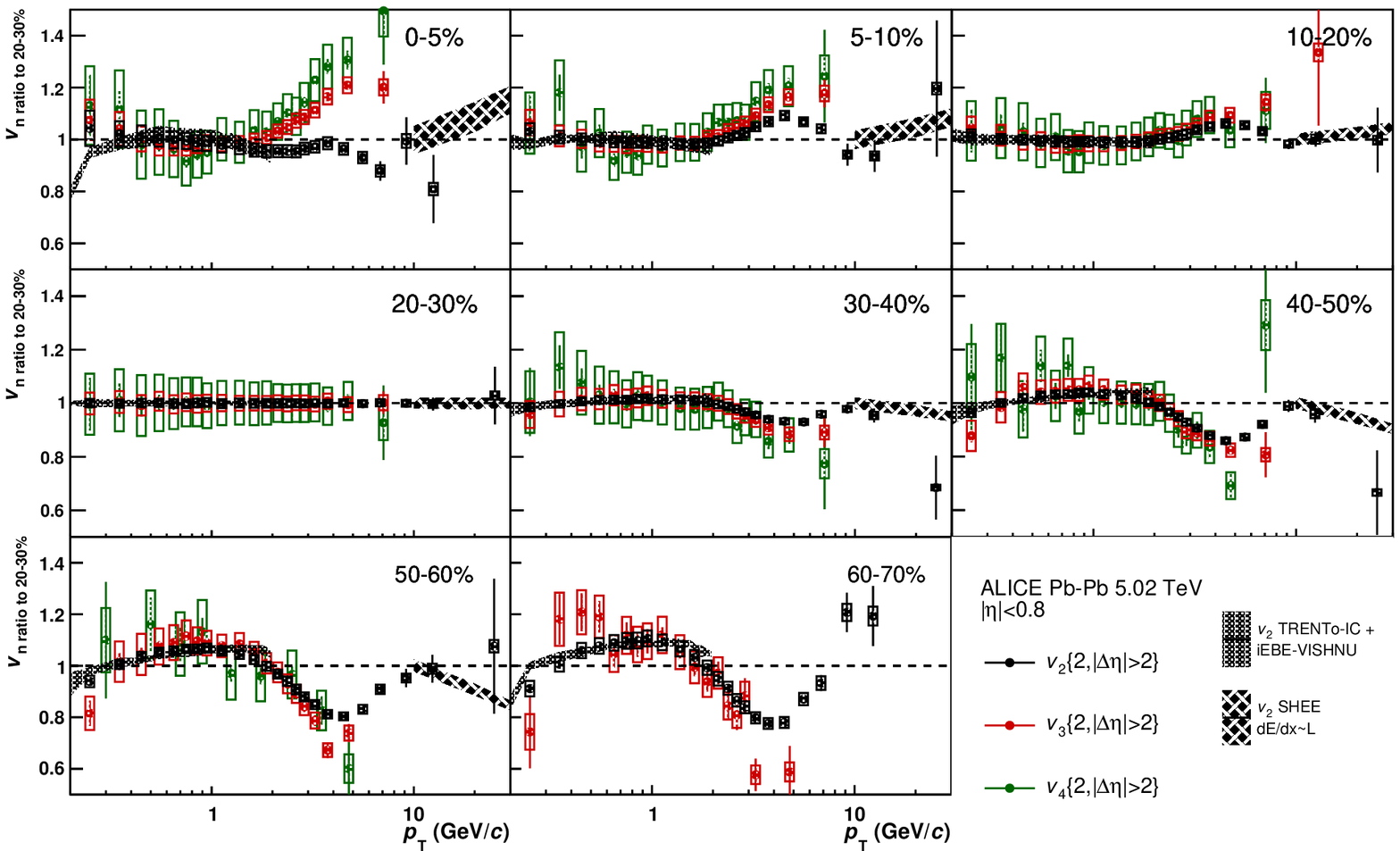}
  \caption{Ratios $ v_n(p_\text{T})_\text{ratio to \text{20-30\%}}$ of inclusive charged particles for Pb--Pb collisions at $\sqrt{{\textit s}_\text{NN}} = 5.02$ TeV, in different centrality classes, measured with the scalar product method with respect to the V0A $Q$-vector. Hydrodynamic calculations \cite{Zhao:2017yhj, Betz:2016ayq} are shown for comparison.}
  \label{fig:shapeev}
\end{figure}

The evolution of the shape of $p_\text{T}$-differential $v_n$ coefficients with respect to centrality is investigated by calculating the ratios of $v_n(p_\text{T})$ in a given centrality range and $v_n(p_\text{T})$ in centrality 20--30\%, normalised by the corresponding ratio of $p_\text{T}$-integrated $v_n$
\begin{equation}
v_n(p_\text{T})_\text{ratio to \text{20-30\%}} = \frac{v_n(p_\text{T})}{v_n(p_\text{T})[\text{20-30\%}]} \frac{v_{n}[\text{20-30\%}]}{v_n} .
\end{equation}
In order to reduce statistical fluctuations, a parametrisation of $v_n(p_\text{T})[\text{20-30\%}]$ fitted to data is employed. If the shape of $v_n(p_\text{T})$ does not change with centrality, $v_n(p_\text{T})_\text{ratio to \text{20-30\%}}$ is identical to 1 in the full $p_\text{T}$ range. The results are shown in Fig.\ \ref{fig:shapeev}: deviations from unity up to about $10\%$ are observed at low $p_\text{T}$ ($p_\text{T}<3$ GeV/$c$) and up to about $30\%$ at intermediate $p_\text{T}$ ($3<p_\text{T}<6$ GeV/$c$), where $v_n(p_\text{T})$ reaches its maximum. These variations are observed to be larger for higher harmonics ($v_{3-4}$), in particular for central collisions. The effects due to a change in particle composition of the inclusive charged particle sample with centrality are estimated to be negligible. These deviations are attributed mostly to the combined effect of radial flow and parton density which, in the coalescence model picture \cite{Fries:2008hs}, decrease from central to peripheral collision shifting the maximum of $v_n(p_\text{T})$ from higher to lower $p_\text{T}$. At high $p_\text{T}$ ($p_\text{T}>10$ GeV/$c$), results on $v_{2}\{2,|\Delta\eta|>2\}$ are consistent with those at low $p_\text{T}$, suggesting a common origin of the centrality evolution of elliptic flow in the two regimes, presumably initial-state geometry and its fluctuations. This interpretation is consistent with the findings of \cite{Sirunyan:2017pan}. The attribution of the scaling of $v_n(p_\text{T})$ up to $p_\text{T} = 8$ GeV/$c$ to initial-state geometries agrees with studies \cite{Aad:2013xma, Adam:2015eta} using the Event Shape Engineering technique \cite{Schukraft:2012ah} and $p_\text{T}$-dependent elliptic flow fluctuations \cite{Abelev:2012di}. Finally, the models using hydrodynamic calculations \cite{Zhao:2017yhj} and jet energy loss ones \cite{Betz:2016ayq} are observed to be in good agreement with the $v_2$ data at low and high $p_\text{T}$, respectively.

\begin{figure}
\centering
  \includegraphics[width=\linewidth]{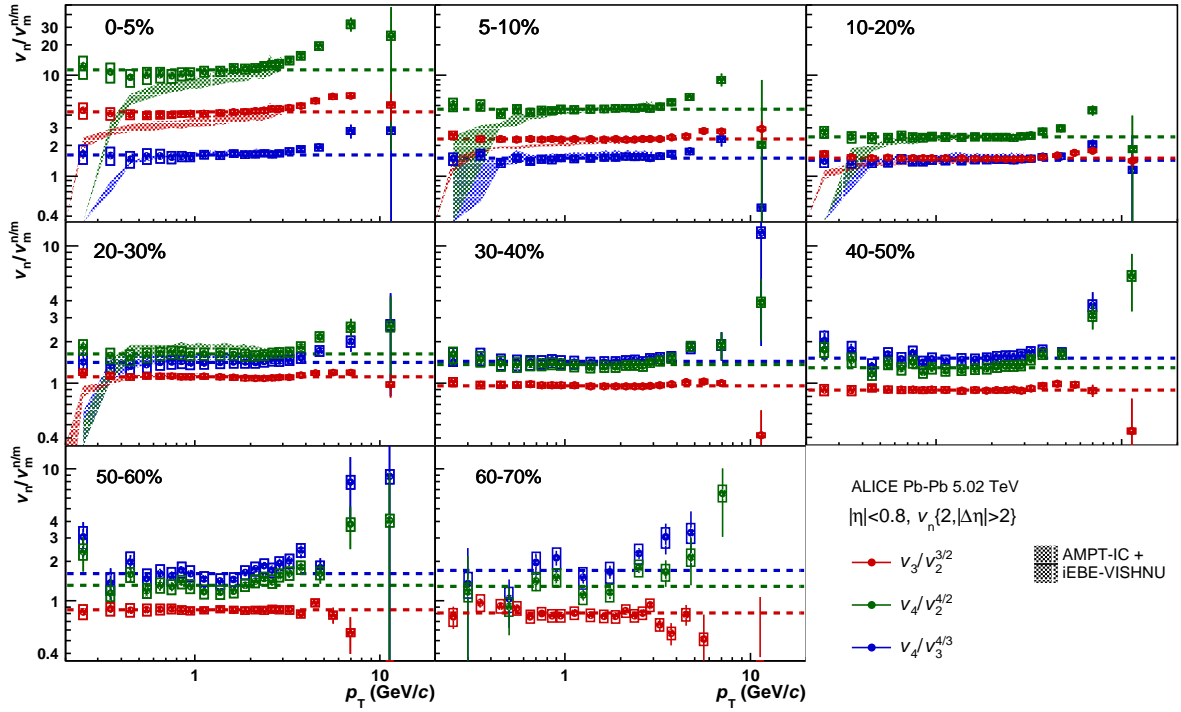}
  \caption{Ratios $v_n(p_\text{T})/v_m(p_\text{T})^{n/m}, \; n=3,4, \; m=2,3$ of inclusive charged particles for Pb--Pb collisions at $\sqrt{{\textit s}_\text{NN}} = 5.02$, in different centrality classes, measured with the scalar product method with respect to the V0A $Q$-vector. Dashed lines represent averages in $0.2 < p_\text{T} < 3$ GeV/$c$. The ratios are also shown for one hydrodynamic model \cite{Zhao:2017yhj} in the four most central centrality intervals; it is qualitatively similar in the other centrality intervals and for the other models.}
  \label{fig:ptscaling}
\end{figure}

At RHIC \cite{Adams:2003zg, Adare:2010ux} and LHC \cite{ATLAS:2012at} it had been observed that the ratios of harmonics follow a power-law scaling, i.e.\ $v_{n}^{1/n} \sim v_{m}^{1/m}$, for semi-central and peripheral collisions up to about $6$ GeV/$c$ and independent of the harmonic $n$ and $m$. In order to test this scaling, we use the ratios $v_{n}/v_{m}^{n/m}$ which in practice are more sensitive than $v_{n}^{1/n} \sim v_{m}^{1/m}$. Figure \ref{fig:ptscaling} shows these ratios for $n = 3,4$ and $m = 2,3$, as a function of $p_\text{T}$. These ratios are indeed observed to be independent of $p_\text{T}$, in most of the $p_\text{T}$ range and for most centrality ranges, except for centrality 0--5\%. Up to about the maximum of $v_{n}(p_\text{T})$, the scaling is numerically related to, but actually significantly more precise than, the observed approximate power-law dependences $v_n(p_\text{T}) \sim p_\text{T}^{n/3}$ pointed out in Fig.\ \ref{fig:7}. Surprisingly however, the scaling extends much further, in particular $v_{3}(p_\text{T})/v_{2}(p_\text{T})^{3/2}$ is constant to better than about 10\%, out to the highest measured $p_\text{T}$ in excess of 10 GeV/$c$. The ratio $v_{4}(p_\text{T})/v_{2}(p_\text{T})^{4/2}$ shows stronger deviations at high $p_\text{T}$, starting at around the maximum of $v_{2}(p_\text{T})$. A separation of $v_{4}$ into linear and non-linear components would be required to see if the $v_{4}/v_{2}$ scaling at low $p_\text{T}$, and/or its violations at high $p_\text{T}$, is related to the mode mixing, which is particularly strong for the 4th harmonic and at high $p_\text{T}$, or possibly also to quark coalescence \cite{Borghini:2005kd, Gombeaud:2009ye, Kolb:2004gi}. 

As noted in the context of Fig.\ \ref{fig:7}, the observed ratio scaling is not expected in  ideal hydrodynamics. While not all viscous hydrodynamical models shown in Fig.\ \ref{fig:hydro} describe the data up to the highest $p_\text{T}$ very well, they all do exhibit the same power-law scaling in the ratio of harmonics over the $p_\text{T}$ range $0.5 < p_\text{T} < 3$ GeV/$c$, with a precision comparable to the one seen in the data, while they strongly deviate for $p_\text{T} < 0.5$ GeV/$c$. 
The scaling may be related to viscosity, as also postulated in \cite{Liu:2018hjh, Lacey:2011ug}, in particular to the large and $p_\text{T}$-dependent viscous corrections appearing at hadronisation \cite{Teaney:2003kp}.
However, a harmonic number dependence of these viscous corrections which could reproduce the scaling observed in the data, has so far, to the best of our knowledge, never been quantitatively investigated.

\section{Elliptic flow fluctuations} \label{sec:flowfluct}

\begin{figure}
  \centering
  \includegraphics[width=0.65\linewidth]{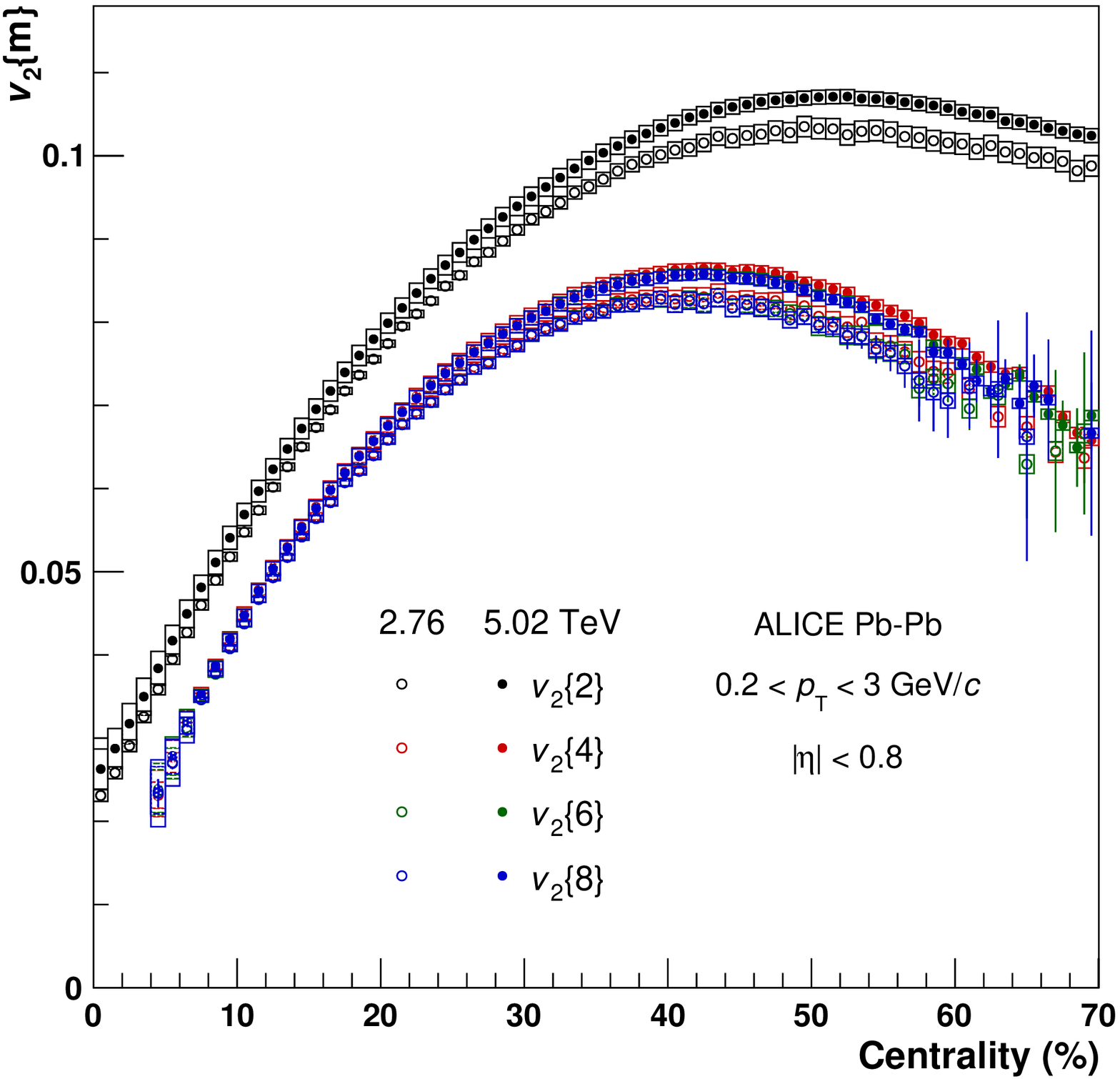}
  \caption{Elliptic flow coefficient $v_2$ of inclusive charged particles as a function of centrality, measured with the two- and multi-particle cumulant methods. Measurements for Pb--Pb collisions at $\sqrt{{\textit s}_\text{NN}} = 5.02$ (2.76) TeV are shown by solid (open) markers.}
  \label{fig:3}
\end{figure}

\begin{figure}
  \centering
  \includegraphics[width=\linewidth]{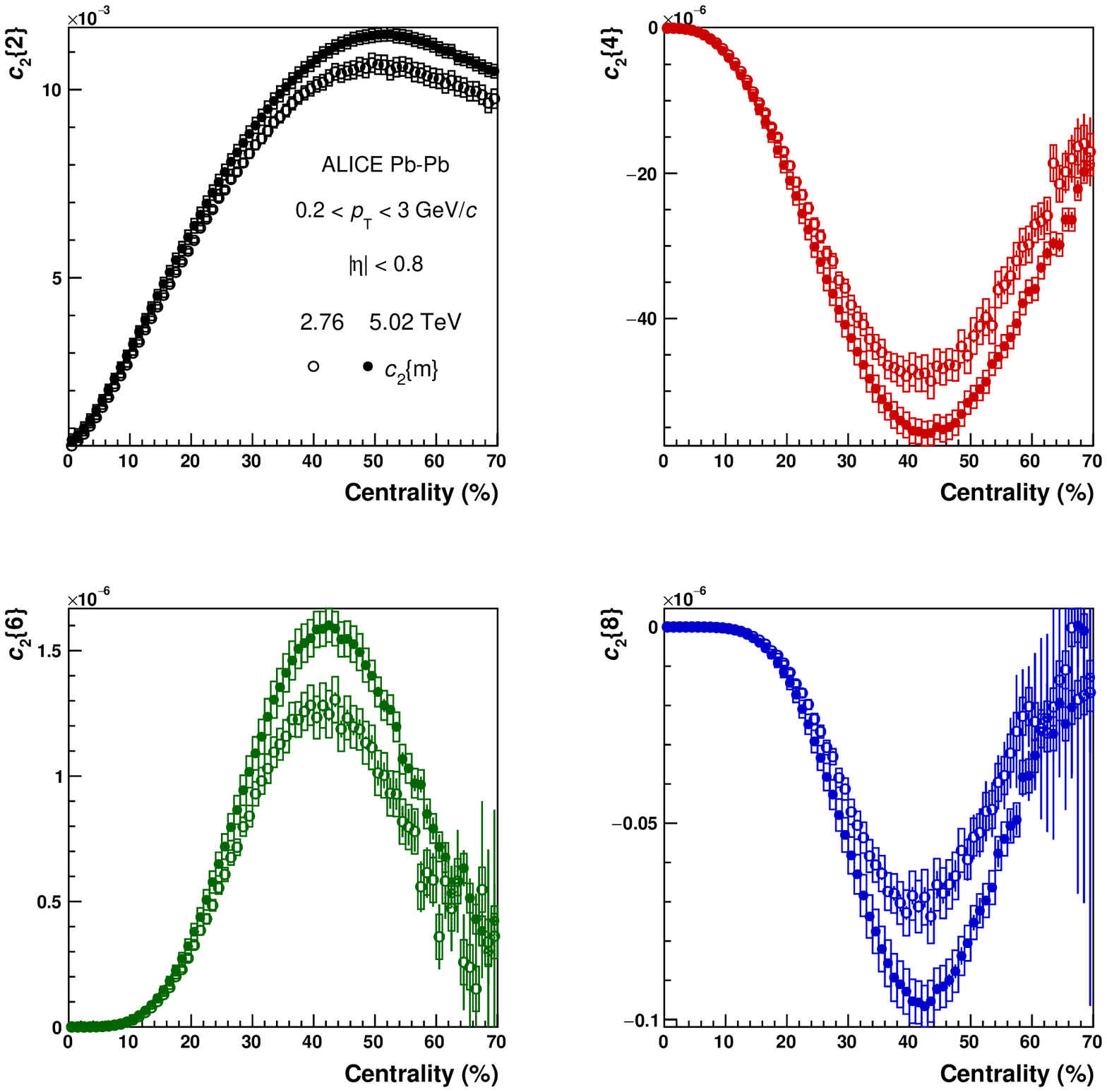}
  \caption{Cumulants $c_2$ of elliptic flow of inclusive charged particles as a function of centrality, for the two-particle and multi-particle cumulant methods. Measurements for Pb--Pb collisions at $\sqrt{{\textit s}_\text{NN}} = 5.02$ (2.76) TeV are shown by solid (open) markers.}
  \label{fig:rawcum}
\end{figure}

Figure \ref{fig:3} shows the integrated $v_2$ in the $p_\text{T}$ range $0.2 < p_\text{T} < 3$ GeV/$c$ as a function of centrality, measured with two-, four-, six- and eight-particle cumulants at $\sqrt{{\textit s}_\text{NN}} = 5.02$ and 2.76 TeV. The corresponding cumulants ($c_2\{2,4,6,8\}$) are reported in Fig.\ \ref{fig:rawcum}. The centrality dependence is similar for all multi-particle cumulants and similar to what is shown in Fig.\ \ref{fig:1}. The differences between $v_2\{2\}$ (shown in Fig.\ \ref{fig:3}) and $v_2\{2,|\Delta\eta|>1\}$ (shown in Fig.\ \ref{fig:1}), which range from about 4\% in mid-central collisions to about 20\% in peripheral ones, are mostly attributed to non-flow contributions, which are suppressed in the case of results with a pseudorapidity gap. The possible differences arising from the decorrelation of event planes at different pseudorapidities are expected to be less than 1\%, as previously argued.

\begin{figure}
  \centering
  \includegraphics[width=0.65\linewidth]{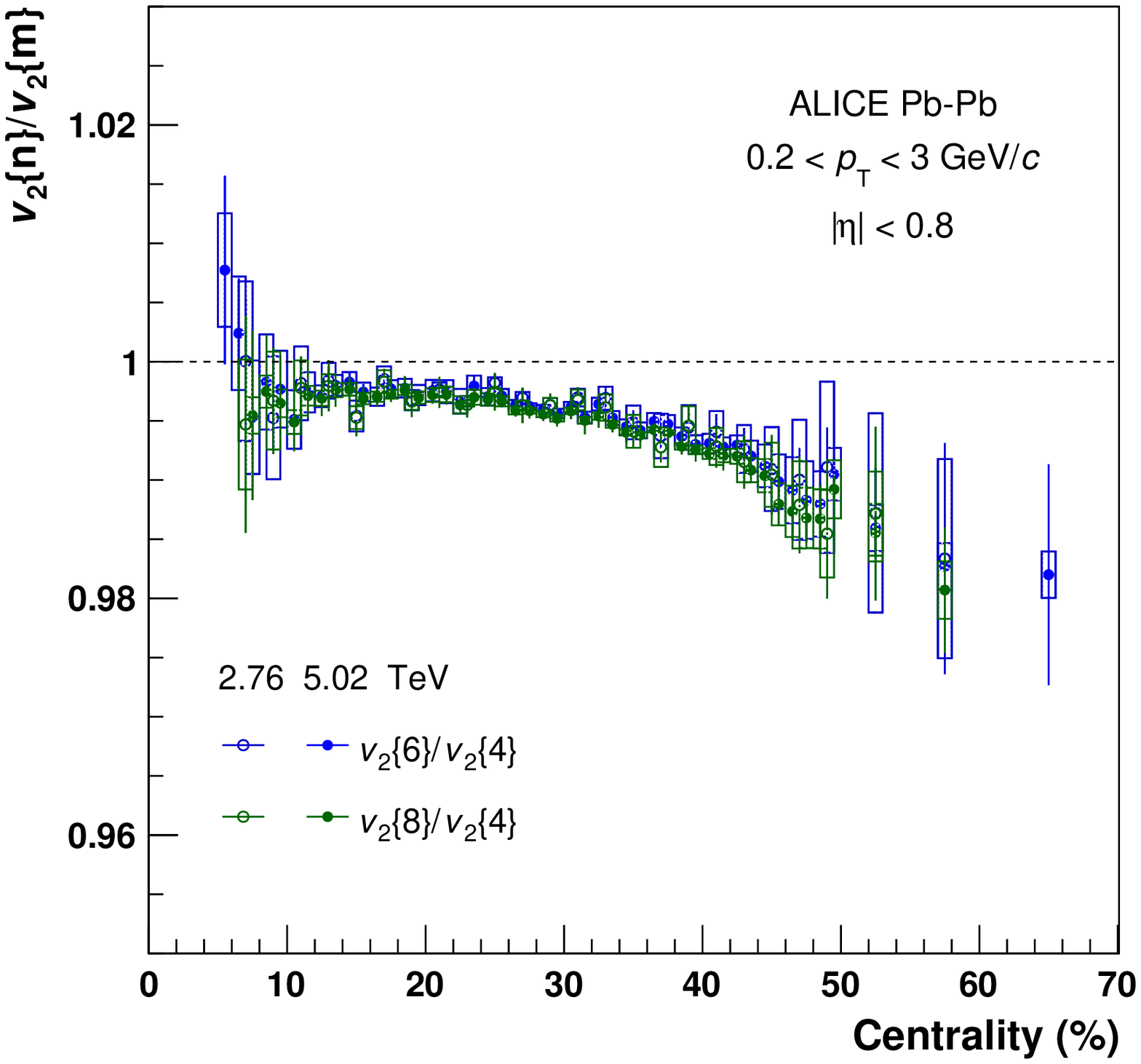}
  \caption{Ratios of elliptic flow coefficients $v_2$ of inclusive charged particles between measurements with different multi-particle cumulant methods, as a function of centrality. Measurements at $\sqrt{{\textit s}_\text{NN}} = 5.02$ (2.76) TeV are shown by solid (open) markers.}
  \label{fig:4}
\end{figure}

\begin{figure}
  \centering
  \includegraphics[width=0.65\linewidth]{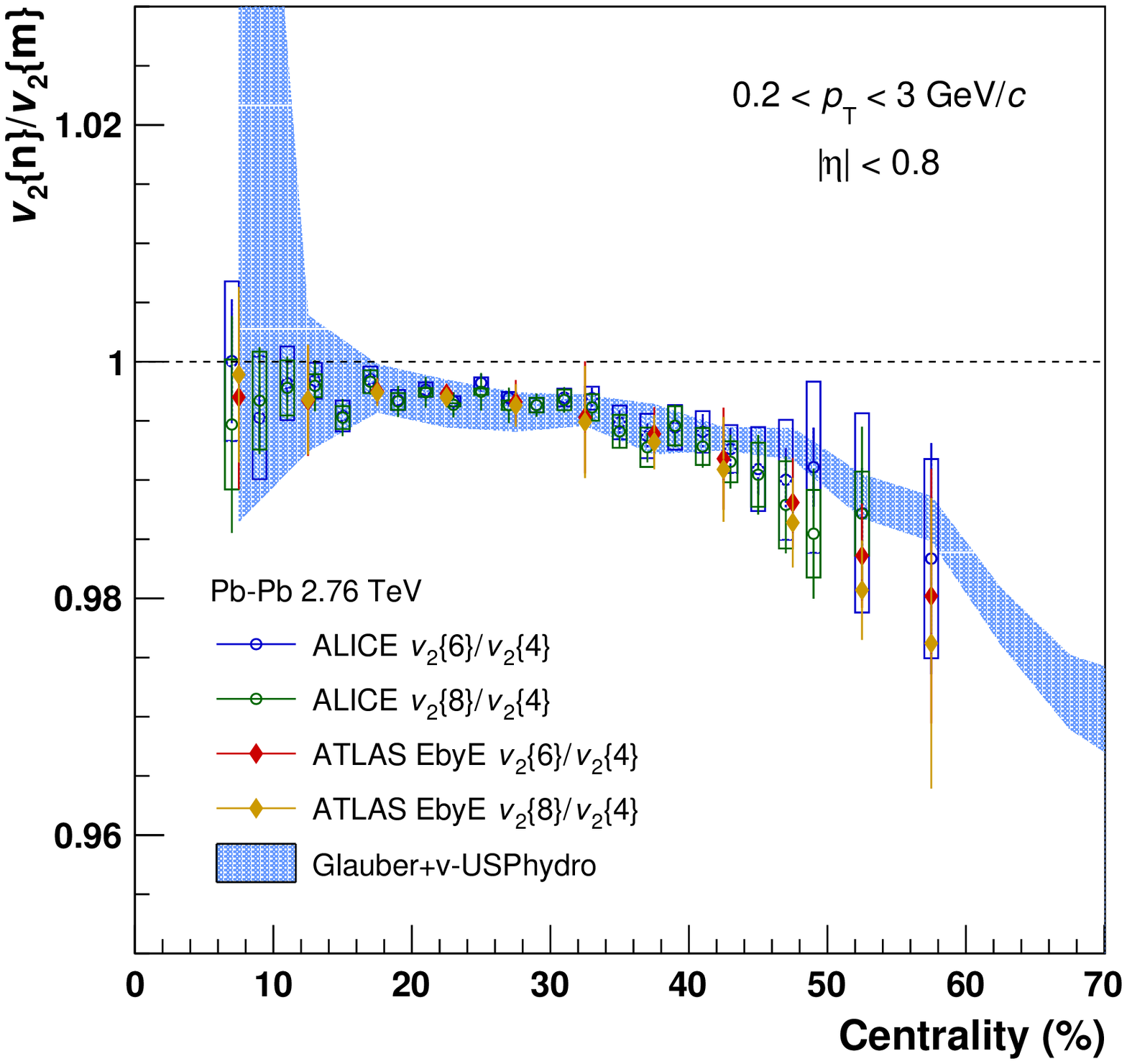}
  \caption{Ratios of elliptic flow coefficients $v_2$ of inclusive charged particles between measurements with different multi-particle cumulant methods, as a function of centrality, at $\sqrt{{\textit s}_\text{NN}} = 2.76$  TeV. Hydrodynamic calculations \cite{Giacalone:2016eyu} and ATLAS measurements \cite{Aad:2014vba} are shown for comparison.}
  \label{fig:ATLAS}
\end{figure}

A fine-splitting of less than 1\% is observed among $v_2\{4\}$, $v_2\{6\}$ and $v_2\{8\}$, as it can be seen from their ratios, shown in Fig.\ \ref{fig:4} for both collision energies. The ratios $v_2\{6\}/v_2\{4\}$ and $v_2\{8\}/v_2\{4\}$ at $\sqrt{{\textit s}_\text{NN}} = 5.02$ TeV show a significant centrality dependence: the deviations of the ratios from unity is about 0.2\% in central and increases up to about 1\% for mid-central collisions. A further increase seems to be observed for more peripheral collisions, up to about 2\% for centralities above 50\%. The systematic uncertainties on these ratios are greately reduced with respect to those on $v_2\{m\}$ $(m=2,4,6,8)$, since the dominant sources of systematic uncertainty (track quality variables and centrality determination) among the two- and multi-particle cumulants are highly correlated. This fine-splitting is consistent with non-Bessel-Gaussian behaviour of event-by-event flow fluctuations, as previously explained. These ratios are found to be independent of the choice of $p_\text{T}$ range within $0.2 < p_\text{T} < 3$ GeV/$c$, indicating that the characterization of flow fluctuations at low $p_\text{T}$ does not depend on $p_\text{T}$, even for such fine-splitting. Results at $\sqrt{{\textit s}_\text{NN}} = 2.76$ TeV are found to be compatible, indicating that these ratios do not change significantly across collision energies. Compared to calculations \cite{Giacalone:2016eyu} employing MC-Glauber initial conditions \cite{Alver:2008aq} and viscous hydrodynamics (v-USPhydro) for Pb--Pb collisions at $\sqrt{{\textit s}_\text{NN}} = 2.76$ TeV, the ratios $v_2\{6\}/v_2\{4\}$ and $v_2\{8\}/v_2\{4\}$ are found to be compatible. A good agreement is found between the results at $\sqrt{{\textit s}_\text{NN}} = 2.76$ TeV and corresponding ATLAS results on elliptic flow p.d.f.\ obtained via the unfolding technique \cite{Aad:2014vba}, as shown in Fig.\ \ref{fig:ATLAS}.

\begin{figure}
    \centering
  \includegraphics[width=0.65\linewidth]{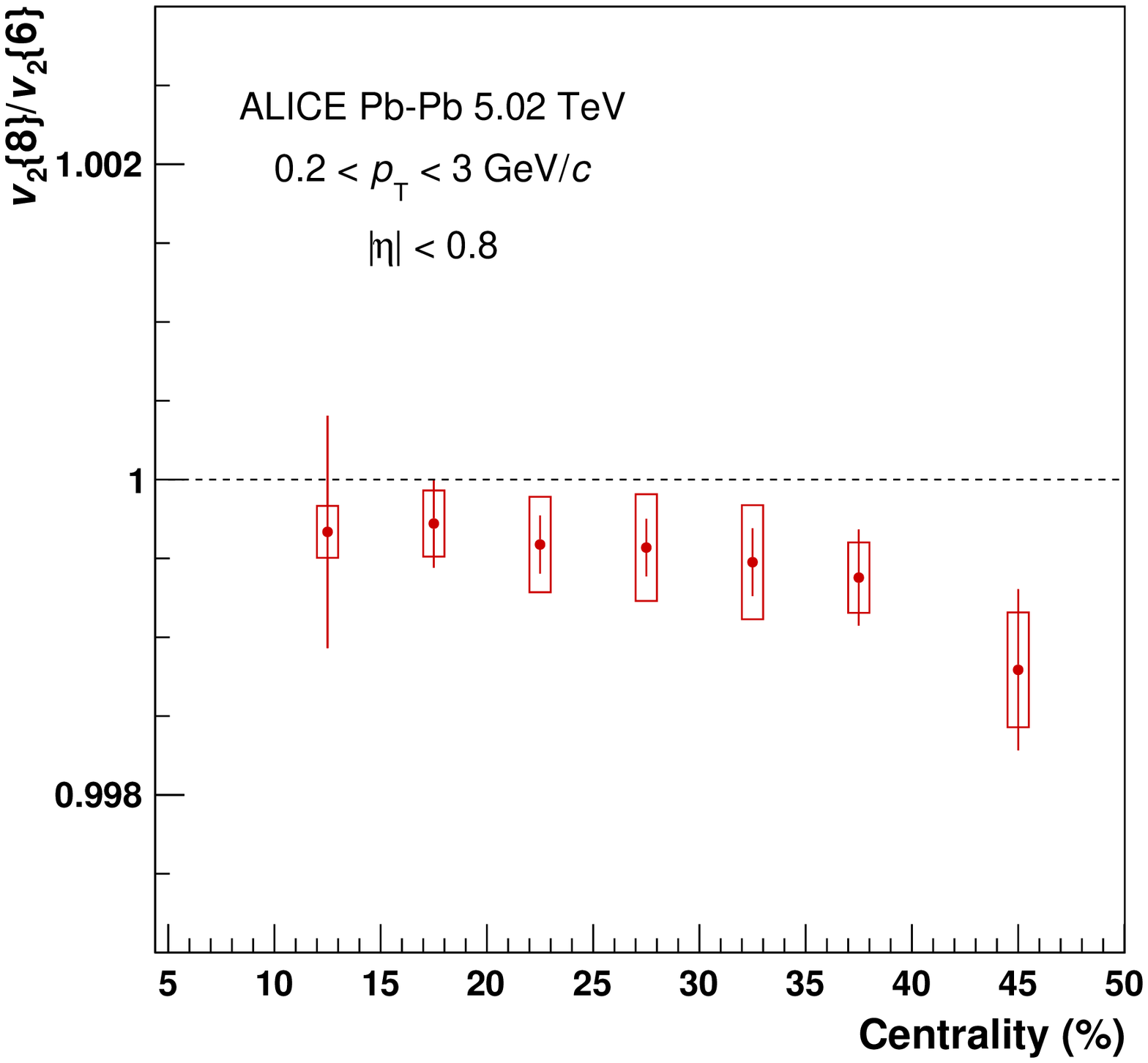}
  \caption{Ratio of elliptic flow coefficients $v_2\{8\}/v_2\{6\}$ of inclusive charged particles as a function of centrality.}
  \label{fig:ratiov8v6}
\end{figure}

\begin{figure}
  \centering
  \includegraphics[width=0.65\linewidth]{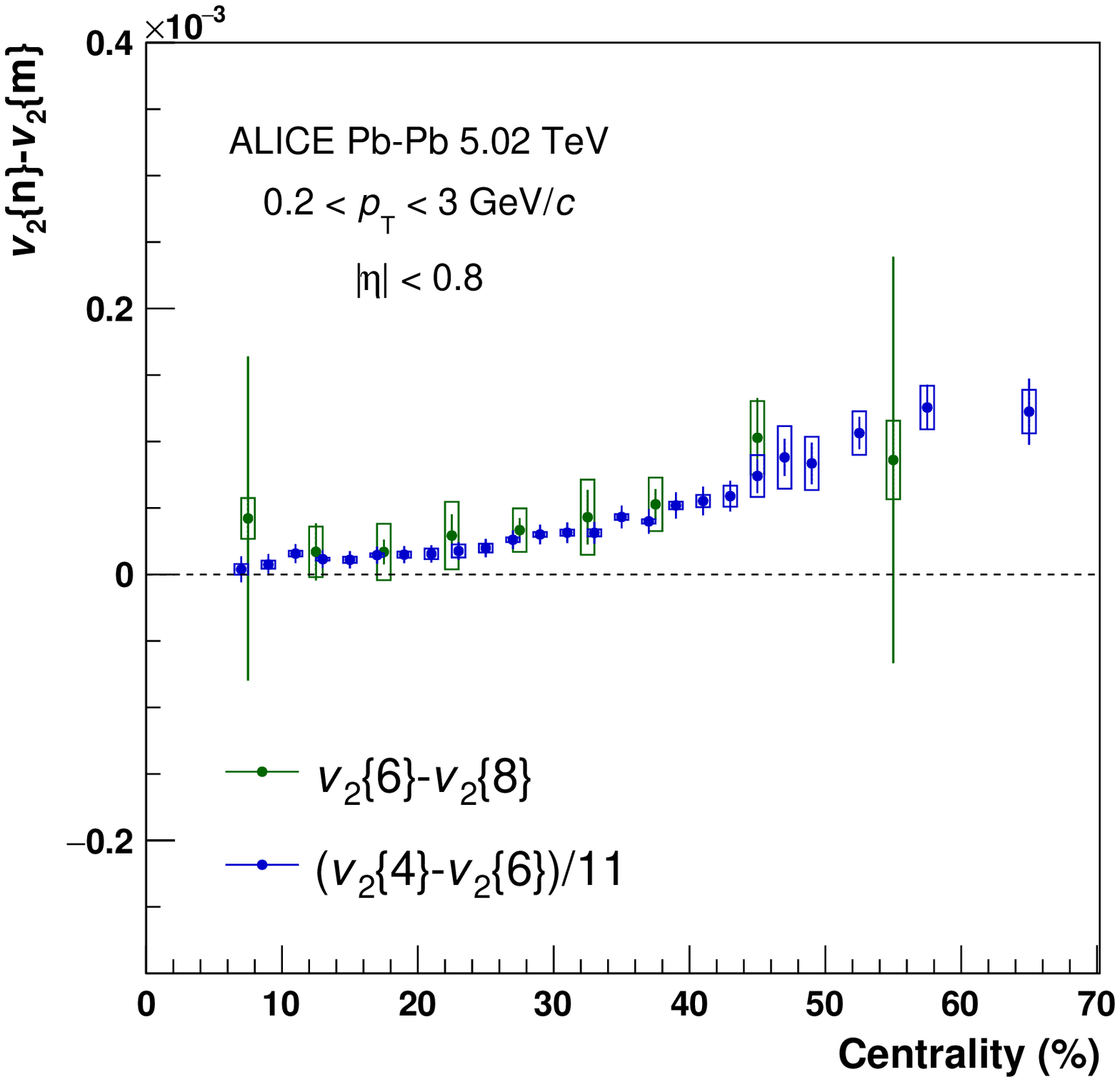}
  \caption{Differences of elliptic flow coefficients $v_2$ of inclusive charged particles between measurements with different multi-particle cumulant methods, as a function of centrality.}
  \label{fig:test}
\end{figure}

Figure \ref{fig:ratiov8v6} shows the ratio between $v_2\{8\}$ and $v_2\{6\}$ at $\sqrt{{\textit s}_\text{NN}} = 5.02$ TeV. A hint of a further fine-splitting between these two, of the order of $0.05$\%, is observed. The results suggest little or no centrality dependence within centrality 10--50\%. This difference is also consistent with non-Bessel-Gaussian elliptic flow fluctuations, and can be attributed to different contributions of the skewness to these higher-order cumulants \cite{Giacalone:2016eyu}. Corresponding results at $\sqrt{{\textit s}_\text{NN}} = 2.76$ TeV, here and in the following, are not shown because of the large statistical uncertainties. Figure \ref{fig:test} shows $v_2\{6\}-v_2\{8\}$ and $(v_2\{4\}-v_2\{6\})/11$ at $\sqrt{{\textit s}_\text{NN}} = 5.02$ TeV: these two are observed to be in agreement, which demonstrates the validity of Eq.\ \ref{eq:1}. This observation sets an upper limit of $4 \times 10^{-4}$ at 95\% confidence level for possible contributions to multi-particle cumulants from higher moments of the flow p.d.f.\ (kurtosis and beyond) in the centrality range 10--50\%. This estimate is obtained assuming gaussian systematic uncertainties and summing them in quadrature with the statistical ones.

\begin{figure}
  \centering
  \includegraphics[width=0.65\linewidth]{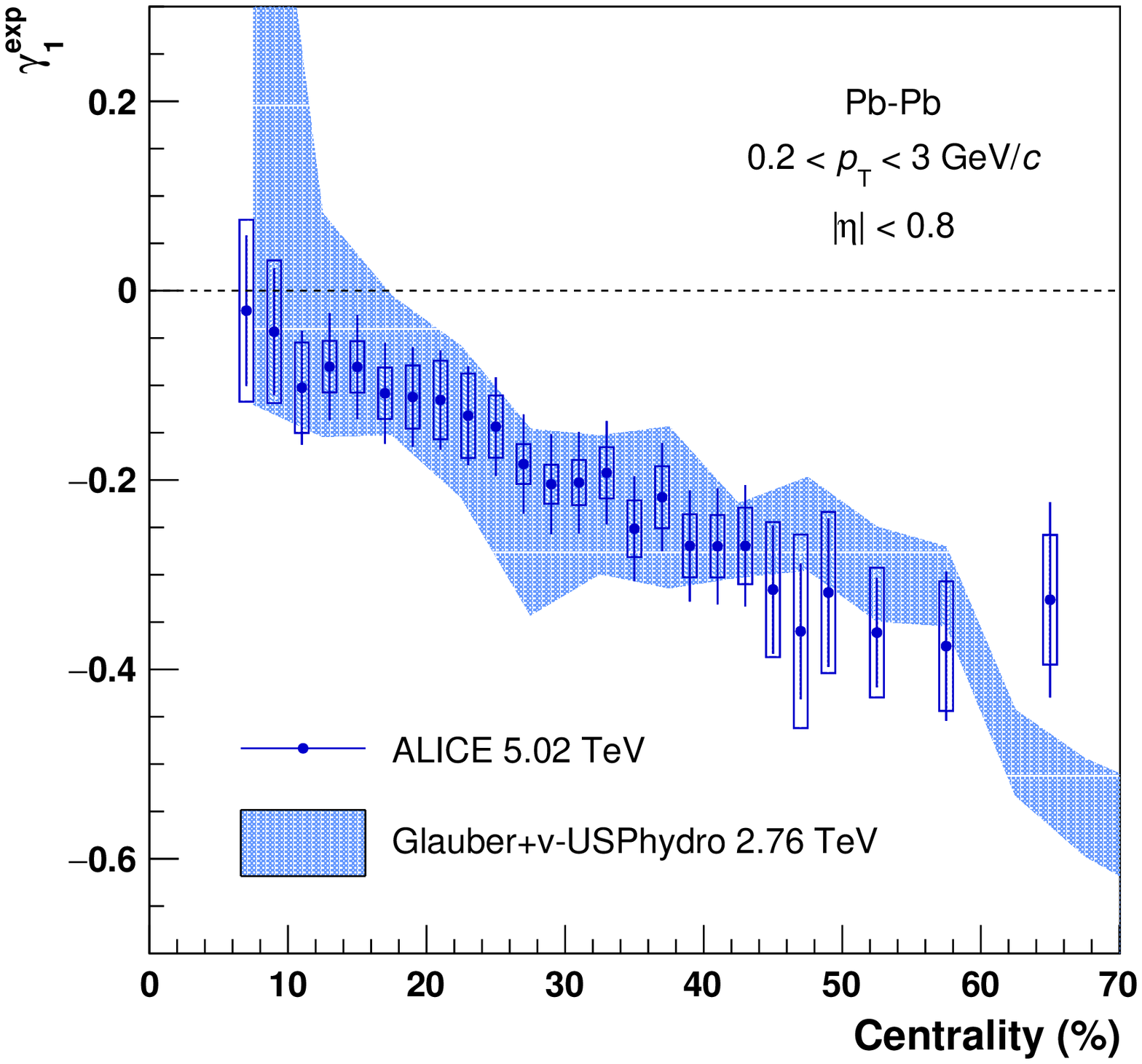}
  \caption{Skewness of elliptic flow $\gamma_1^\text{exp}$ of inclusive charged particles as a function of centrality, for Pb--Pb collisions at $\sqrt{{\textit s}_\text{NN}} = 5.02$ TeV. Hydrodynamic calculations \cite{Giacalone:2016eyu} for Pb--Pb collisions at $\sqrt{{\textit s}_\text{NN}} = 2.76$ TeV are shown for comparison.}
  \label{fig:skew}
\end{figure}

Figure \ref{fig:skew} shows the measurement of the standardised skewness ($\gamma_1^\text{exp}$) at $\sqrt{{\textit s}_\text{NN}} = 5.02$ TeV as a function of centrality. To suppress non-flow contributions, the values of $v_2\{2,|\Delta\eta|>1\}$ from Fig.\ \ref{fig:1} are used for $v_2\{2\}$ in Eq.\ \ref{eq:gamma}.
A negative value of the skewness, with a strong centrality dependence, is observed: $\gamma_1^\text{exp}$ decreases from zero in central to about $- 0.4$ in peripheral collisions. Compared to model calculations \cite{Giacalone:2016eyu} for Pb--Pb collisions at $\sqrt{{\textit s}_\text{NN}} = 2.76$ TeV, the results are found to be compatible for the entire centrality range. This observation is consistent with the elliptic flow p.d.f. being progressively more left-skewed going from central to peripheral collisions. We attribute this feature to the combination of an increase in $\langle \varepsilon_2 \rangle$ and the geometrical constrain $\varepsilon_2 < 1$, as previously argued.

\begin{figure}
\centering
  \includegraphics[width=\linewidth]{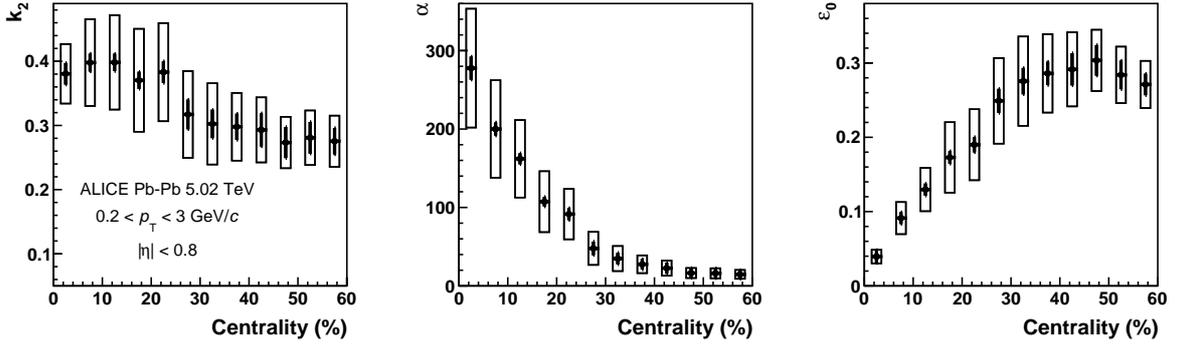}
  \caption{Elliptic power parameters $k_2$, $\alpha$ and $\varepsilon_0$ as a function of centrality, for Pb--Pb collisions at $\sqrt{{\textit s}_\text{NN}} = 5.02$ TeV, extracted from measurements of $v_2$ of inclusive charged particles with two-particle and multi-particle cumulant methods.}
  \label{fig:EPpara}
\end{figure}

In order to report the full p.d.f.\ of elliptic flow $P(v_2)$, which can be compared to previous experimental results and theoretical predictions, it is parametrised with the Elliptic Power distribution \cite{Yan:2013laa, Yan:2014afa}
\begin{equation} \label{eq:EP}
P(v_2) = \frac{\text{d}\varepsilon_2}{\text{d}v_2} \, P(\varepsilon_2) = \frac{1}{k_2} \, P\left(\frac{v_2}{k_2}\right) = \frac{2 \alpha v_2}{\pi k_2^2} \, ( 1 - \varepsilon_{0}^2)^{\alpha+1/2} \, \int_0^\pi \frac{(1 - v_2^2/k_2^2)^{\alpha-1}}{(1 - v_{2} \varepsilon_{0} \cos\varphi / k_2)^{2\alpha+1}} \, \text{d}\varphi,
\end{equation}
and its three free parameters ($\alpha$, $\varepsilon_0$ and $k_2$) are extracted from fits to the elliptic flow cumulants $c_2\{2,|\Delta\eta|>1\}$ and $c_2\{m\}$ $(m=4,6,8)$ at $\sqrt{{\textit s}_\text{NN}} = 5.02$ TeV. The parameter $\alpha$ quantifies the magnitude of elliptic flow fluctuations, $\varepsilon_0$ the mean eccentricity in the reaction plane and $k_2$ is the proportionality coefficient between initial-state eccentricity and $v_2$ coefficient: $v_2 = k_2 \varepsilon_2$. The relation between cumulants and Elliptic Power parameters is given by \cite{Yan:2014afa}
\begin{align}
c_2\{2\} = &k_2^2 \, \left( 1 - f_1 \right), \label{eq:EPfit1} \\
c_2\{4\} = &- k_2^4 \, \left( 1 - 2 \, f_1  + 2 \, f_1^2 - f_2 \right), \label{eq:EPfit2} \\
c_2\{6\} = &k_2^6 \, \left( 4 + 18 \, f_1^2 - 12 \, f_1^3 + 12 f_1 \left( 3 f_2 - 1 \right) - 6 \, f_2 - \, f_3 \right), \label{eq:EPfit3} \\ 
c_2\{8\} = &- k_2^8 \, ( 33 - 288 \, f_1^3 + 144 \, f_1^4 - 66 \, f_2 + 18 \, f_2^2 - 24 \, f_1^2 (-11 + 6 \, f_2) \nonumber \\ 
&- 12 \, f_3 + 4 \, f_1 ( -33 + 42 \, f_2 + 4 \, f_3 ) - f_4 ) \, \label{eq:EPfit4}
\end{align}
where
\begin{equation}
f_k \equiv \langle ( 1 - \varepsilon_n^2)^k \rangle = \frac{\alpha}{\alpha+k} \, (1-\varepsilon_0^2)^k \, _2 F_1 \left( k+ \frac{1}{2}, k; \alpha+k+1, \varepsilon_0^2\right)
\end{equation}
and $_2F_1$ is the hypergeometric function. The results are shown in Fig.\ \ref{fig:EPpara}. The systematic uncertainties are assigned varying the fit ranges and initial values of the parameters and shifting the data points according to the corresponding systematic uncertainties. An additional source of uncertainty, which is investigated, is a possible cubic response coefficient $k_2'$, defined as $v_2 = k_2 \varepsilon_2 + k_2' \varepsilon_2^3$. This coefficient is introduced to quantify the possible increase of flow fluctuations that the hydrodynamic expansion of the medium introduces with respect to geometrical fluctuations in the initial state and was argued to be non-zero in mid-central and peripheral collisions due to general properties of the hydrodynamic phase \cite{Noronha-Hostler:2015dbi}. In particular, $k_2'$ is expected to be $\leq 0.15$ in the centrality range 0--60\% \cite{Noronha-Hostler:2015dbi}. The residual differences in $\alpha$, $\varepsilon_0$ and $k_2$ when including $k_2'$ as an additional free parameter are considered in the systematic uncertainties. The statistical uncertainties are evaluated using the subsampling method: the analysed dataset is divided into 10 sub-samples and $v_2\{m\}$ is measured in each of them. The Elliptic Power parameters are then extracted in each subsample and their dispersion is used to estimate the statistical uncertainties.

\begin{figure}
\centering
  \includegraphics[width=\linewidth]{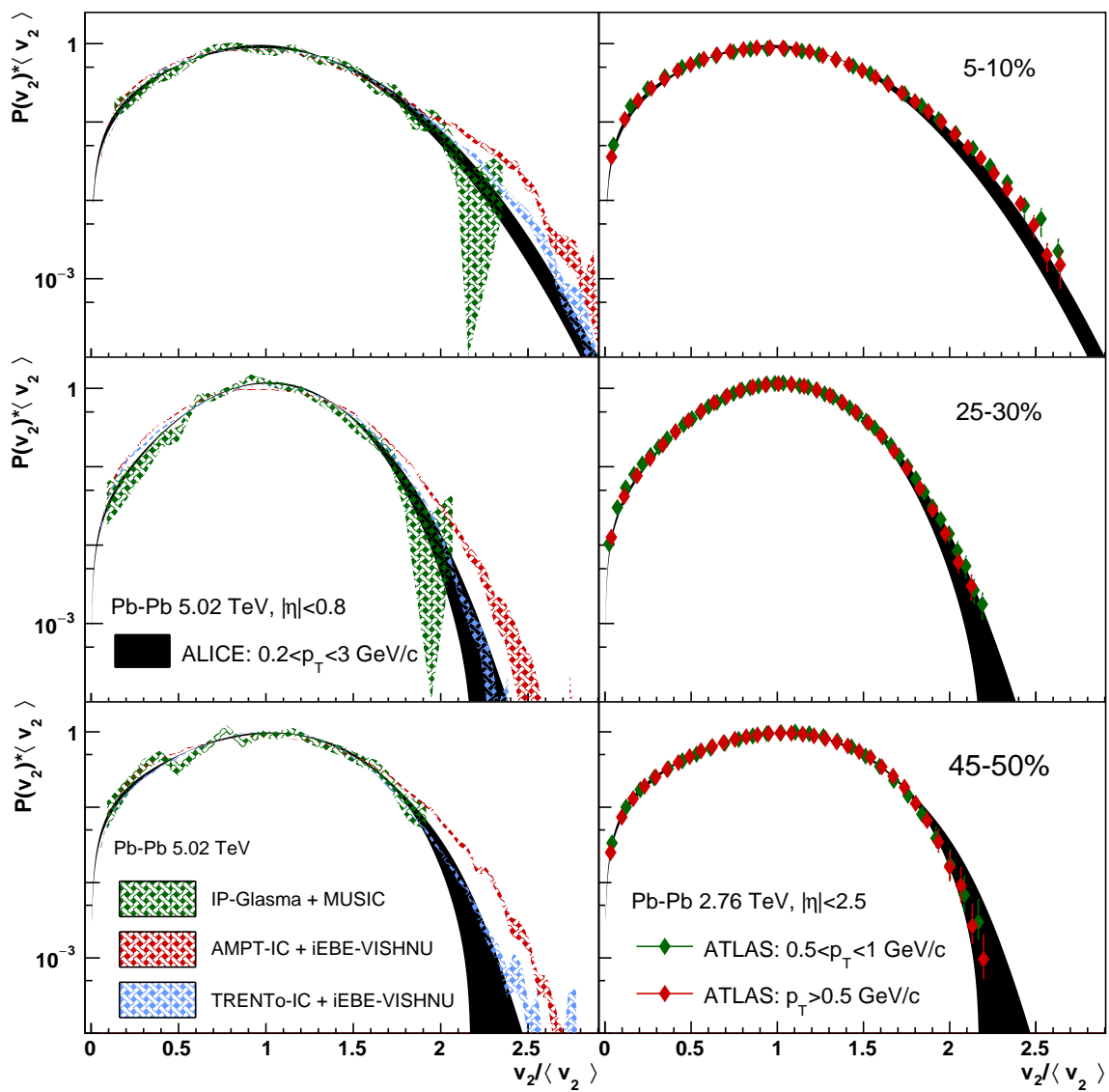}
  \label{fig:5}
  \caption{Elliptic flow p.d.f. $P(v_2)$ rescaled by the mean $v_2$ ($\langle v_2 \rangle$) of inclusive charged particles for Pb--Pb collisions at $\sqrt{{\textit s}_\text{NN}} = 5.02$ TeV, in different centrality classes. Several hydrodynamic calculations \cite{McDonald:2016vlt, Zhao:2017yhj} and previous measurements from ATLAS \cite{Aad:2013xma} at lower energies are shown for comparison.}
  \label{fig:flowdist}
\end{figure}

The resulting p.d.f., constructed using the Elliptic Power distribution (Eq.\ \ref{eq:EP}) with the parameters shown in Fig.\ \ref{fig:EPpara} and scaled by its mean ($\langle v_2 \rangle$), is reported in Fig.\ \ref{fig:flowdist}, for centralities 5--10\%, 25--30\% and 45--50\%. The systematic uncertainties take into account the correlation of the uncertainties of the Elliptic Power parameters. Other centrality ranges that are not shown here are reported in the appendix \ref{appA}. Scaling by $\langle v_2 \rangle$ allows a comparison of our data with results by the ATLAS collaboration \cite{Aad:2013xma} obtained in different $p_\text{T}$ ranges. The observed agreement is also consistent, as previously noted, with elliptic flow fluctuations at low $p_\text{T}$ not depending on $p_\text{T}$ and not changing significantly between collision energies, except for the trivial increase in $p_\text{T}$-integrated $v_2$ due to the change in $\langle p_\text{T} \rangle$. Comparison with iEBE-VISHNU model calculations with AMPT and TRENTo initial conditions \cite{Zhao:2017yhj} indicates that TRENTo initial conditions are better at describing the experimental data. The data are found to be in agreement also with predictions from the IP-Glasma+MUSIC model \cite{McDonald:2016vlt}  (with initial conditions very similar to the TRENTo ones), although the uncertainties on the theoretical predictions do not allow to draw firm conclusions.

\section{Conclusions}

Anisotropic flow coefficients are measured up to the sixth harmonic for inclusive charged particles at mid-rapidity ($|\eta|<0.8$), in a wide centrality (0--80\%) and $p_\text{T}$ ($0.2 < p_\text{T} < 50$ GeV/$c$) ranges, for Pb--Pb collisions at $\sqrt{{\textit s}_\text{NN}}$ = 5.02 and 2.76 TeV. Comparing the results at $\sqrt{{\textit s}_\text{NN}}$ = 5.02 and 2.76 TeV the energy dependence of anisotropic flow at the LHC is investigated. Comparison with different model calculations demonstrates that these measurements have the potential to constrain initial-state fluctuations, transport parameters of the medium and path-length dependence of energy loss of high-$p_\text{T}$ partons. The evolution of $v_n(p_\text{T})$ with respect to centrality and harmonic number $n$ is also investigated. Flow coefficient of all harmonics are observed to follow an approximate power-law scaling of the form $v_n(p_\text{T}) \sim p_\text{T}^{n/3}$ in the $p_\text{T}$ range $0.2 < p_\text{T} < 3$ GeV/$c$. The ratios $v_n/v_m^{n/m}$ $n = 3,4$ and $m = 2,3$ are also observed to be independent of $p_\text{T}$ within the same $p_\text{T}$ range and show deviations of about 10\% for $3 < p_\text{T} < 10$ GeV/$c$.

The fluctuations of elliptic flow are investigated through the fine-splitting of the higher-order multi-particle cumulants ($v_2\{4\}$, $v_2\{6\}$, $v_2\{8\}$), from which the standardised skewness ($\gamma_1^\text{exp}$) of the flow p.d.f.\ is extracted. Results are found to be compatible both with predictions from hydrodynamical models and with previous ATLAS results at lower energies. It is concluded that the characterization of elliptic flow  fluctuations at low $p_\text{T}$ does not depend on the $p_\text{T}$ range and on the collision energy, except for the increase in $p_\text{T}$-integrated $v_2$ due to the change in $\langle p_\text{T} \rangle$. Direct constraints on the contribution of higher moments to the multi-particle cumulants are also reported. Finally, the full elliptic flow p.d.f., parametrised with the Elliptic Power distribution, is reported in the centrality ranges 0--60\%. These results are also found to be in agreement with previous experimental results. Overall, calculations including initial conditions matching the IP-Glasma description are observed to better reproduce the elliptic flow p.d.f.\ while failing to describe the $p_\text{T}$ dependence of anisotropic flow coefficients, whereas the opposite situation is observed for calculations that employ AMPT initial conditions.

\newenvironment{acknowledgement}{\relax}{\relax}
\begin{acknowledgement}
\section*{Acknowledgements}

The ALICE Collaboration would like to thank all its engineers and technicians for their invaluable contributions to the construction of the experiment and the CERN accelerator teams for the outstanding performance of the LHC complex.
The ALICE Collaboration gratefully acknowledges the resources and support provided by all Grid centres and the Worldwide LHC Computing Grid (WLCG) collaboration.
The ALICE Collaboration acknowledges the following funding agencies for their support in building and running the ALICE detector:
A. I. Alikhanyan National Science Laboratory (Yerevan Physics Institute) Foundation (ANSL), State Committee of Science and World Federation of Scientists (WFS), Armenia;
Austrian Academy of Sciences and Nationalstiftung f\"{u}r Forschung, Technologie und Entwicklung, Austria;
Ministry of Communications and High Technologies, National Nuclear Research Center, Azerbaijan;
Conselho Nacional de Desenvolvimento Cient\'{\i}fico e Tecnol\'{o}gico (CNPq), Universidade Federal do Rio Grande do Sul (UFRGS), Financiadora de Estudos e Projetos (Finep) and Funda\c{c}\~{a}o de Amparo \`{a} Pesquisa do Estado de S\~{a}o Paulo (FAPESP), Brazil;
Ministry of Science \& Technology of China (MSTC), National Natural Science Foundation of China (NSFC) and Ministry of Education of China (MOEC) , China;
Ministry of Science and Education, Croatia;
Ministry of Education, Youth and Sports of the Czech Republic, Czech Republic;
The Danish Council for Independent Research | Natural Sciences, the Carlsberg Foundation and Danish National Research Foundation (DNRF), Denmark;
Helsinki Institute of Physics (HIP), Finland;
Commissariat \`{a} l'Energie Atomique (CEA) and Institut National de Physique Nucl\'{e}aire et de Physique des Particules (IN2P3) and Centre National de la Recherche Scientifique (CNRS), France;
Bundesministerium f\"{u}r Bildung, Wissenschaft, Forschung und Technologie (BMBF) and GSI Helmholtzzentrum f\"{u}r Schwerionenforschung GmbH, Germany;
General Secretariat for Research and Technology, Ministry of Education, Research and Religions, Greece;
National Research, Development and Innovation Office, Hungary;
Department of Atomic Energy Government of India (DAE), Department of Science and Technology, Government of India (DST), University Grants Commission, Government of India (UGC) and Council of Scientific and Industrial Research (CSIR), India;
Indonesian Institute of Science, Indonesia;
Centro Fermi - Museo Storico della Fisica e Centro Studi e Ricerche Enrico Fermi and Istituto Nazionale di Fisica Nucleare (INFN), Italy;
Institute for Innovative Science and Technology , Nagasaki Institute of Applied Science (IIST), Japan Society for the Promotion of Science (JSPS) KAKENHI and Japanese Ministry of Education, Culture, Sports, Science and Technology (MEXT), Japan;
Consejo Nacional de Ciencia (CONACYT) y Tecnolog\'{i}a, through Fondo de Cooperaci\'{o}n Internacional en Ciencia y Tecnolog\'{i}a (FONCICYT) and Direcci\'{o}n General de Asuntos del Personal Academico (DGAPA), Mexico;
Nederlandse Organisatie voor Wetenschappelijk Onderzoek (NWO), Netherlands;
The Research Council of Norway, Norway;
Commission on Science and Technology for Sustainable Development in the South (COMSATS), Pakistan;
Pontificia Universidad Cat\'{o}lica del Per\'{u}, Peru;
Ministry of Science and Higher Education and National Science Centre, Poland;
Korea Institute of Science and Technology Information and National Research Foundation of Korea (NRF), Republic of Korea;
Ministry of Education and Scientific Research, Institute of Atomic Physics and Romanian National Agency for Science, Technology and Innovation, Romania;
Joint Institute for Nuclear Research (JINR), Ministry of Education and Science of the Russian Federation and National Research Centre Kurchatov Institute, Russia;
Ministry of Education, Science, Research and Sport of the Slovak Republic, Slovakia;
National Research Foundation of South Africa, South Africa;
Centro de Aplicaciones Tecnol\'{o}gicas y Desarrollo Nuclear (CEADEN), Cubaenerg\'{\i}a, Cuba and Centro de Investigaciones Energ\'{e}ticas, Medioambientales y Tecnol\'{o}gicas (CIEMAT), Spain;
Swedish Research Council (VR) and Knut \& Alice Wallenberg Foundation (KAW), Sweden;
European Organization for Nuclear Research, Switzerland;
National Science and Technology Development Agency (NSDTA), Suranaree University of Technology (SUT) and Office of the Higher Education Commission under NRU project of Thailand, Thailand;
Turkish Atomic Energy Agency (TAEK), Turkey;
National Academy of  Sciences of Ukraine, Ukraine;
Science and Technology Facilities Council (STFC), United Kingdom;
National Science Foundation of the United States of America (NSF) and United States Department of Energy, Office of Nuclear Physics (DOE NP), United States of America.
\end{acknowledgement}

\bibliographystyle{utphys}
\bibliography{paper.bib}


\newpage
\appendix

\section{Additional figures} \label{appA}

\counterwithin{figure}{section}

\begin{figure}[H]
\centering
  \includegraphics[width=\linewidth]{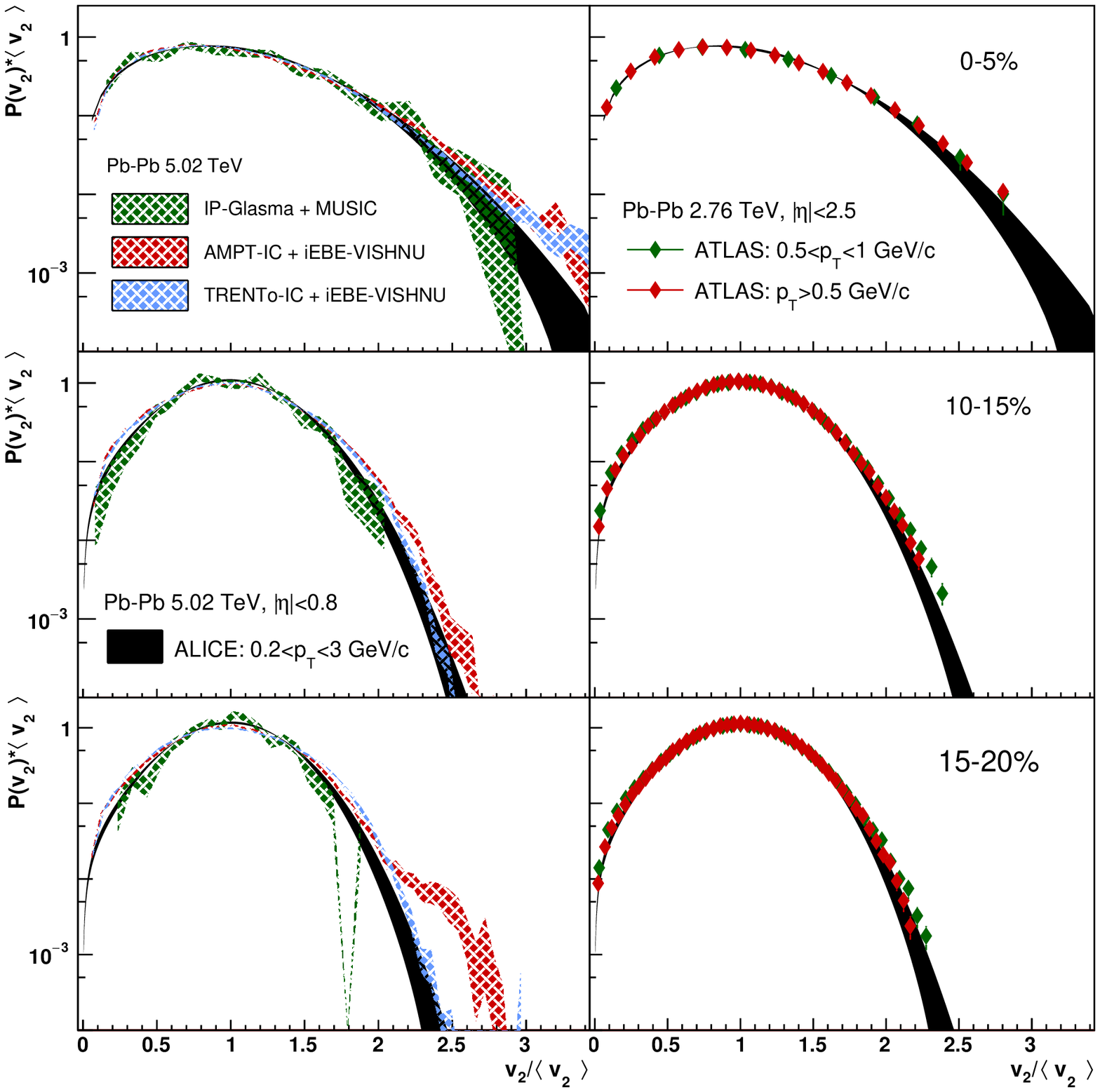}
  \caption{Elliptic flow p.d.f. $P(v_2)$ rescaled by $\langle v_2 \rangle$ in centralities 0--5\%, 10--15\% and 15--20\% for Pb--Pb collisions at $\sqrt{{\textit s}_\text{NN}} = 5.02$ TeV.}
  \label{fig:flowdist2}
\end{figure}

\begin{figure}
\centering
  \includegraphics[width=\linewidth]{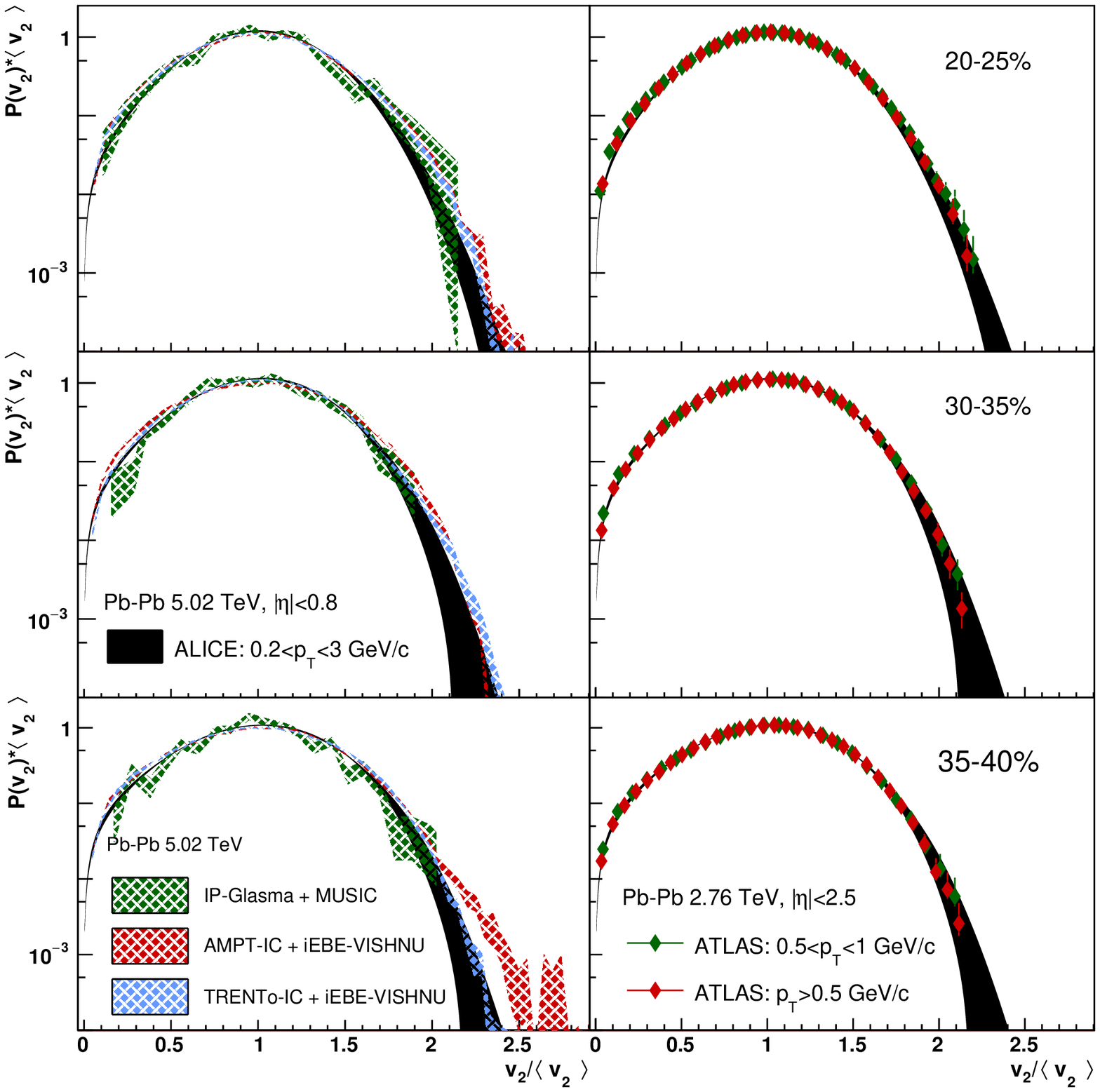}
  \caption{Elliptic flow p.d.f. $P(v_2)$ rescaled by $\langle v_2 \rangle$ in centralities 20--25\%, 30--35\% and 35--40\% for Pb--Pb collisions at $\sqrt{{\textit s}_\text{NN}} = 5.02$ TeV.}
  \label{fig:flowdist3}
\end{figure}

\begin{figure}
\centering
  \includegraphics[width=\linewidth]{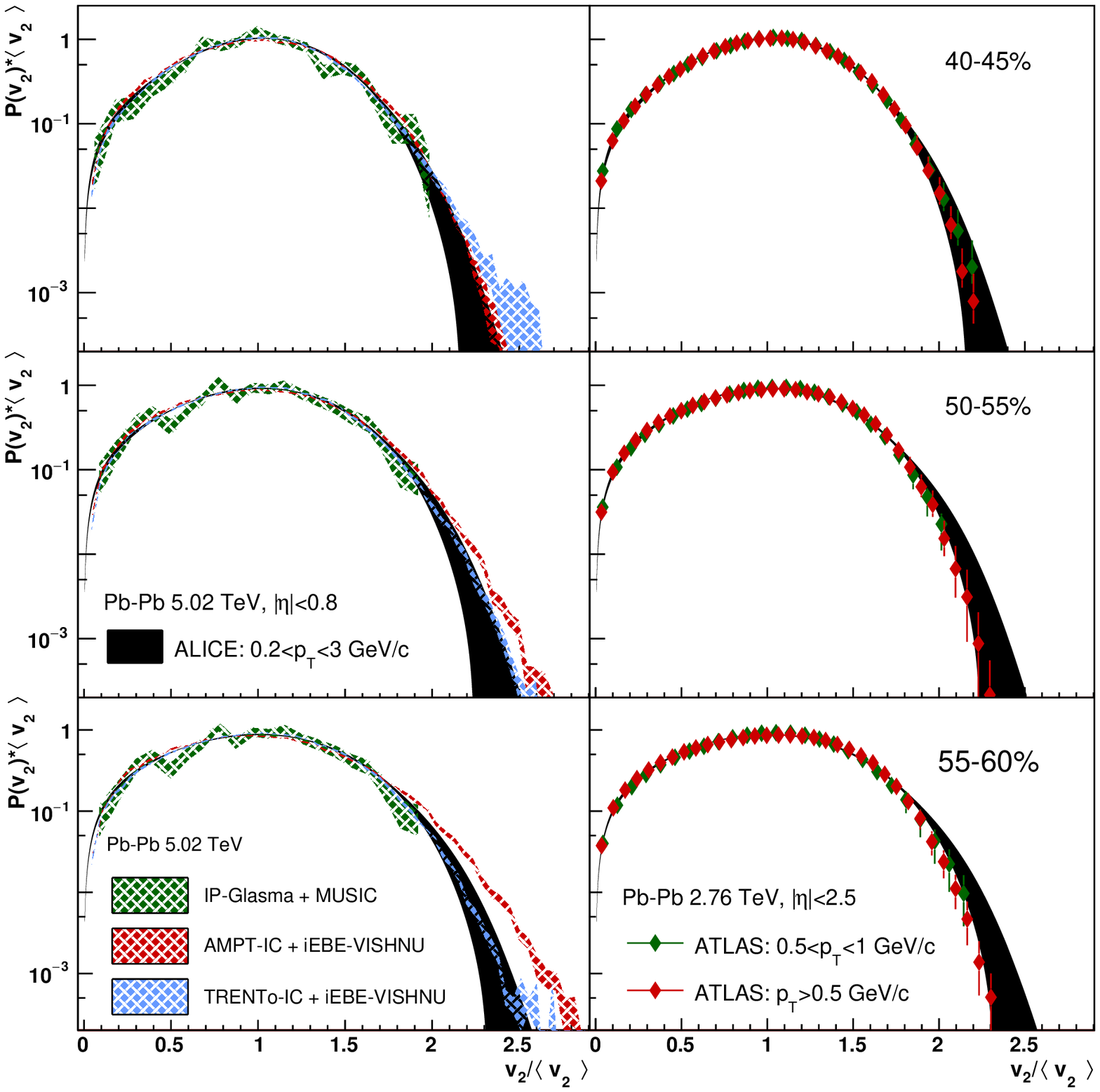}
  \caption{Elliptic flow p.d.f. $P(v_2)$ rescaled by $\langle v_2 \rangle$ in centralities 40--45\%, 50--55\% and 55--60\% for Pb--Pb collisions at $\sqrt{{\textit s}_\text{NN}} = 5.02$ TeV.}
  \label{fig:flowdist4}
\end{figure}

The elliptic flow p.d.f. $P(v_2)$, constructed as explained in Sec.\ \ref{sec:flowfluct}, in the centrality ranges not shown in Fig.\ \ref{fig:flowdist} are reported in Fig. \ref{fig:flowdist2}--\ref{fig:flowdist4}.

\newpage
\section{The ALICE Collaboration}
\label{app:collab}

\begingroup
\small
\begin{flushleft}
S.~Acharya\Irefn{org138}\And 
F.T.-.~Acosta\Irefn{org22}\And 
D.~Adamov\'{a}\Irefn{org93}\And 
J.~Adolfsson\Irefn{org80}\And 
M.M.~Aggarwal\Irefn{org97}\And 
G.~Aglieri Rinella\Irefn{org36}\And 
M.~Agnello\Irefn{org33}\And 
N.~Agrawal\Irefn{org48}\And 
Z.~Ahammed\Irefn{org138}\And 
S.U.~Ahn\Irefn{org76}\And 
S.~Aiola\Irefn{org143}
% ...
\And
\renewcommand\labelenumi{\textsuperscript{\theenumi}~}

\section*{Affiliation notes}
\renewcommand\theenumi{\roman{enumi}}
\begin{Authlist}
\item \Adef{orgI}Dipartimento DET del Politecnico di Torino, Turin, Italy
\item \Adef{orgII}M.V. Lomonosov Moscow State University, D.V. Skobeltsyn Institute of Nuclear, Physics, Moscow, Russia
\item \Adef{orgIII}Department of Applied Physics, Aligarh Muslim University, Aligarh, India
\item \Adef{orgIV}Institute of Theoretical Physics, University of Wroclaw, Poland
\end{Authlist}

\section*{Collaboration Institutes}
\renewcommand\theenumi{\arabic{enumi}~}
\begin{Authlist}
\item \Idef{org1}A.I. Alikhanyan National Science Laboratory (Yerevan Physics Institute) Foundation, Yerevan, Armenia
\item \Idef{org2}Benem\'{e}rita Universidad Aut\'{o}noma de Puebla, Puebla, Mexico
\item \Idef{org3}Bogolyubov Institute for Theoretical Physics, National Academy of Sciences of Ukraine, Kiev, Ukraine
\item \Idef{org4}Bose Institute, Department of Physics  and Centre for Astroparticle Physics and Space Science (CAPSS), Kolkata, India
\item \Idef{org5}Budker Institute for Nuclear Physics, Novosibirsk, Russia
\item \Idef{org6}California Polytechnic State University, San Luis Obispo, California, United States
\item \Idef{org7}Central China Normal University, Wuhan, China
\item \Idef{org8}Centre de Calcul de l'IN2P3, Villeurbanne, Lyon, France
\item \Idef{org9}Centro de Aplicaciones Tecnol\'{o}gicas y Desarrollo Nuclear (CEADEN), Havana, Cuba
\item \Idef{org10}Centro de Investigaci\'{o}n y de Estudios Avanzados (CINVESTAV), Mexico City and M\'{e}rida, Mexico
\item \Idef{org11}Centro Fermi - Museo Storico della Fisica e Centro Studi e Ricerche ``Enrico Fermi', Rome, Italy
\item \Idef{org12}Chicago State University, Chicago, Illinois, United States
\item \Idef{org13}China Institute of Atomic Energy, Beijing, China
\item \Idef{org14}Chonbuk National University, Jeonju, Republic of Korea
\item \Idef{org15}Comenius University Bratislava, Faculty of Mathematics, Physics and Informatics, Bratislava, Slovakia
\item \Idef{org16}COMSATS Institute of Information Technology (CIIT), Islamabad, Pakistan
\item \Idef{org17}Creighton University, Omaha, Nebraska, United States
\item \Idef{org18}Department of Physics, Aligarh Muslim University, Aligarh, India
\item \Idef{org19}Department of Physics, Ohio State University, Columbus, Ohio, United States
\item \Idef{org20}Department of Physics, Pusan National University, Pusan, Republic of Korea
\item \Idef{org21}Department of Physics, Sejong University, Seoul, Republic of Korea
\item \Idef{org22}Department of Physics, University of California, Berkeley, California, United States
\item \Idef{org23}Department of Physics, University of Oslo, Oslo, Norway
\item \Idef{org24}Department of Physics and Technology, University of Bergen, Bergen, Norway
\item \Idef{org25}Dipartimento di Fisica dell'Universit\`{a} 'La Sapienza' and Sezione INFN, Rome, Italy
\item \Idef{org26}Dipartimento di Fisica dell'Universit\`{a} and Sezione INFN, Cagliari, Italy
\item \Idef{org27}Dipartimento di Fisica dell'Universit\`{a} and Sezione INFN, Trieste, Italy
\item \Idef{org28}Dipartimento di Fisica dell'Universit\`{a} and Sezione INFN, Turin, Italy
\item \Idef{org29}Dipartimento di Fisica e Astronomia dell'Universit\`{a} and Sezione INFN, Bologna, Italy
\item \Idef{org30}Dipartimento di Fisica e Astronomia dell'Universit\`{a} and Sezione INFN, Catania, Italy
\item \Idef{org31}Dipartimento di Fisica e Astronomia dell'Universit\`{a} and Sezione INFN, Padova, Italy
\item \Idef{org32}Dipartimento di Fisica `E.R.~Caianiello' dell'Universit\`{a} and Gruppo Collegato INFN, Salerno, Italy
\item \Idef{org33}Dipartimento DISAT del Politecnico and Sezione INFN, Turin, Italy
\item \Idef{org34}Dipartimento di Scienze e Innovazione Tecnologica dell'Universit\`{a} del Piemonte Orientale and INFN Sezione di Torino, Alessandria, Italy
\item \Idef{org35}Dipartimento Interateneo di Fisica `M.~Merlin' and Sezione INFN, Bari, Italy
\item \Idef{org36}European Organization for Nuclear Research (CERN), Geneva, Switzerland
\item \Idef{org37}Faculty of Engineering and Science, Western Norway University of Applied Sciences, Bergen, Norway
\item \Idef{org38}Faculty of Nuclear Sciences and Physical Engineering, Czech Technical University in Prague, Prague, Czech Republic
\item \Idef{org39}Faculty of Science, P.J.~\v{S}af\'{a}rik University, Ko\v{s}ice, Slovakia
\item \Idef{org40}Frankfurt Institute for Advanced Studies, Johann Wolfgang Goethe-Universit\"{a}t Frankfurt, Frankfurt, Germany
\item \Idef{org41}Gangneung-Wonju National University, Gangneung, Republic of Korea
\item \Idef{org42}Gauhati University, Department of Physics, Guwahati, India
\item \Idef{org43}Helmholtz-Institut f\"{u}r Strahlen- und Kernphysik, Rheinische Friedrich-Wilhelms-Universit\"{a}t Bonn, Bonn, Germany
\item \Idef{org44}Helsinki Institute of Physics (HIP), Helsinki, Finland
\item \Idef{org45}Hiroshima University, Hiroshima, Japan
\item \Idef{org46}Hochschule Worms, Zentrum  f\"{u}r Technologietransfer und Telekommunikation (ZTT), Worms, Germany
\item \Idef{org47}Horia Hulubei National Institute of Physics and Nuclear Engineering, Bucharest, Romania
\item \Idef{org48}Indian Institute of Technology Bombay (IIT), Mumbai, India
\item \Idef{org49}Indian Institute of Technology Indore, Indore, India
\item \Idef{org50}Indonesian Institute of Sciences, Jakarta, Indonesia
\item \Idef{org51}INFN, Laboratori Nazionali di Frascati, Frascati, Italy
\item \Idef{org52}INFN, Sezione di Bari, Bari, Italy
\item \Idef{org53}INFN, Sezione di Bologna, Bologna, Italy
\item \Idef{org54}INFN, Sezione di Cagliari, Cagliari, Italy
\item \Idef{org55}INFN, Sezione di Catania, Catania, Italy
\item \Idef{org56}INFN, Sezione di Padova, Padova, Italy
\item \Idef{org57}INFN, Sezione di Roma, Rome, Italy
\item \Idef{org58}INFN, Sezione di Torino, Turin, Italy
\item \Idef{org59}INFN, Sezione di Trieste, Trieste, Italy
\item \Idef{org60}Inha University, Incheon, Republic of Korea
\item \Idef{org61}Institut de Physique Nucl\'{e}aire d'Orsay (IPNO), Institut National de Physique Nucl\'{e}aire et de Physique des Particules (IN2P3/CNRS), Universit\'{e} de Paris-Sud, Universit\'{e} Paris-Saclay, Orsay, France
\item \Idef{org62}Institute for Nuclear Research, Academy of Sciences, Moscow, Russia
\item \Idef{org63}Institute for Subatomic Physics, Utrecht University/Nikhef, Utrecht, Netherlands
\item \Idef{org64}Institute for Theoretical and Experimental Physics, Moscow, Russia
\item \Idef{org65}Institute of Experimental Physics, Slovak Academy of Sciences, Ko\v{s}ice, Slovakia
\item \Idef{org66}Institute of Physics, Bhubaneswar, India
\item \Idef{org67}Institute of Physics of the Czech Academy of Sciences, Prague, Czech Republic
\item \Idef{org68}Institute of Space Science (ISS), Bucharest, Romania
\item \Idef{org69}Institut f\"{u}r Kernphysik, Johann Wolfgang Goethe-Universit\"{a}t Frankfurt, Frankfurt, Germany
\item \Idef{org70}Instituto de Ciencias Nucleares, Universidad Nacional Aut\'{o}noma de M\'{e}xico, Mexico City, Mexico
\item \Idef{org71}Instituto de F\'{i}sica, Universidade Federal do Rio Grande do Sul (UFRGS), Porto Alegre, Brazil
\item \Idef{org72}Instituto de F\'{\i}sica, Universidad Nacional Aut\'{o}noma de M\'{e}xico, Mexico City, Mexico
\item \Idef{org73}iThemba LABS, National Research Foundation, Somerset West, South Africa
\item \Idef{org74}Johann-Wolfgang-Goethe Universit\"{a}t Frankfurt Institut f\"{u}r Informatik, Fachbereich Informatik und Mathematik, Frankfurt, Germany
\item \Idef{org75}Joint Institute for Nuclear Research (JINR), Dubna, Russia
\item \Idef{org76}Korea Institute of Science and Technology Information, Daejeon, Republic of Korea
\item \Idef{org77}KTO Karatay University, Konya, Turkey
\item \Idef{org78}Laboratoire de Physique Subatomique et de Cosmologie, Universit\'{e} Grenoble-Alpes, CNRS-IN2P3, Grenoble, France
\item \Idef{org79}Lawrence Berkeley National Laboratory, Berkeley, California, United States
\item \Idef{org80}Lund University Department of Physics, Division of Particle Physics, Lund, Sweden
\item \Idef{org81}Nagasaki Institute of Applied Science, Nagasaki, Japan
\item \Idef{org82}Nara Women{'}s University (NWU), Nara, Japan
\item \Idef{org83}National and Kapodistrian University of Athens, School of Science, Department of Physics , Athens, Greece
\item \Idef{org84}National Centre for Nuclear Research, Warsaw, Poland
\item \Idef{org85}National Institute of Science Education and Research, HBNI, Jatni, India
\item \Idef{org86}National Nuclear Research Center, Baku, Azerbaijan
\item \Idef{org87}National Research Centre Kurchatov Institute, Moscow, Russia
\item \Idef{org88}Niels Bohr Institute, University of Copenhagen, Copenhagen, Denmark
\item \Idef{org89}Nikhef, National institute for subatomic physics, Amsterdam, Netherlands
\item \Idef{org90}NRC ¿Kurchatov Institute¿ ¿ IHEP , Protvino, Russia
\item \Idef{org91}NRNU Moscow Engineering Physics Institute, Moscow, Russia
\item \Idef{org92}Nuclear Physics Group, STFC Daresbury Laboratory, Daresbury, United Kingdom
\item \Idef{org93}Nuclear Physics Institute of the Czech Academy of Sciences, \v{R}e\v{z} u Prahy, Czech Republic
\item \Idef{org94}Oak Ridge National Laboratory, Oak Ridge, Tennessee, United States
\item \Idef{org95}Petersburg Nuclear Physics Institute, Gatchina, Russia
\item \Idef{org96}Physics department, Faculty of science, University of Zagreb, Zagreb, Croatia
\item \Idef{org97}Physics Department, Panjab University, Chandigarh, India
\item \Idef{org98}Physics Department, University of Jammu, Jammu, India
\item \Idef{org99}Physics Department, University of Rajasthan, Jaipur, India
\item \Idef{org100}Physikalisches Institut, Eberhard-Karls-Universit\"{a}t T\"{u}bingen, T\"{u}bingen, Germany
\item \Idef{org101}Physikalisches Institut, Ruprecht-Karls-Universit\"{a}t Heidelberg, Heidelberg, Germany
\item \Idef{org102}Physik Department, Technische Universit\"{a}t M\"{u}nchen, Munich, Germany
\item \Idef{org103}Research Division and ExtreMe Matter Institute EMMI, GSI Helmholtzzentrum f\"ur Schwerionenforschung GmbH, Darmstadt, Germany
\item \Idef{org104}Rudjer Bo\v{s}kovi\'{c} Institute, Zagreb, Croatia
\item \Idef{org105}Russian Federal Nuclear Center (VNIIEF), Sarov, Russia
\item \Idef{org106}Saha Institute of Nuclear Physics, Kolkata, India
\item \Idef{org107}School of Physics and Astronomy, University of Birmingham, Birmingham, United Kingdom
\item \Idef{org108}Secci\'{o}n F\'{\i}sica, Departamento de Ciencias, Pontificia Universidad Cat\'{o}lica del Per\'{u}, Lima, Peru
\item \Idef{org109}Shanghai Institute of Applied Physics, Shanghai, China
\item \Idef{org110}Stefan Meyer Institut f\"{u}r Subatomare Physik (SMI), Vienna, Austria
\item \Idef{org111}SUBATECH, IMT Atlantique, Universit\'{e} de Nantes, CNRS-IN2P3, Nantes, France
\item \Idef{org112}Suranaree University of Technology, Nakhon Ratchasima, Thailand
\item \Idef{org113}Technical University of Ko\v{s}ice, Ko\v{s}ice, Slovakia
\item \Idef{org114}Technische Universit\"{a}t M\"{u}nchen, Excellence Cluster 'Universe', Munich, Germany
\item \Idef{org115}The Henryk Niewodniczanski Institute of Nuclear Physics, Polish Academy of Sciences, Cracow, Poland
\item \Idef{org116}The University of Texas at Austin, Austin, Texas, United States
\item \Idef{org117}Universidad Aut\'{o}noma de Sinaloa, Culiac\'{a}n, Mexico
\item \Idef{org118}Universidade de S\~{a}o Paulo (USP), S\~{a}o Paulo, Brazil
\item \Idef{org119}Universidade Estadual de Campinas (UNICAMP), Campinas, Brazil
\item \Idef{org120}Universidade Federal do ABC, Santo Andre, Brazil
\item \Idef{org121}University College of Southeast Norway, Tonsberg, Norway
\item \Idef{org122}University of Cape Town, Cape Town, South Africa
\item \Idef{org123}University of Houston, Houston, Texas, United States
\item \Idef{org124}University of Jyv\"{a}skyl\"{a}, Jyv\"{a}skyl\"{a}, Finland
\item \Idef{org125}University of Liverpool, Department of Physics Oliver Lodge Laboratory , Liverpool, United Kingdom
\item \Idef{org126}University of Split, Faculty of Electrical Engineering, Mechanical Engineering and Naval Architecture, Split, Croatia
\item \Idef{org127}University of Tennessee, Knoxville, Tennessee, United States
\item \Idef{org128}University of the Witwatersrand, Johannesburg, South Africa
\item \Idef{org129}University of Tokyo, Tokyo, Japan
\item \Idef{org130}University of Tsukuba, Tsukuba, Japan
\item \Idef{org131}Universit\'{e} Clermont Auvergne, CNRS/IN2P3, LPC, Clermont-Ferrand, France
\item \Idef{org132}Universit\'{e} de Lyon, Universit\'{e} Lyon 1, CNRS/IN2P3, IPN-Lyon, Villeurbanne, Lyon, France
\item \Idef{org133}Universit\'{e} de Strasbourg, CNRS, IPHC UMR 7178, F-67000 Strasbourg, France, Strasbourg, France
\item \Idef{org134} Universit\'{e} Paris-Saclay Centre d¿\'Etudes de Saclay (CEA), IRFU, Department de Physique Nucl\'{e}aire (DPhN), Saclay, France
\item \Idef{org135}Universit\`{a} degli Studi di Pavia, Pavia, Italy
\item \Idef{org136}Universit\`{a} di Brescia, Brescia, Italy
\item \Idef{org137}V.~Fock Institute for Physics, St. Petersburg State University, St. Petersburg, Russia
\item \Idef{org138}Variable Energy Cyclotron Centre, Kolkata, India
\item \Idef{org139}Warsaw University of Technology, Warsaw, Poland
\item \Idef{org140}Wayne State University, Detroit, Michigan, United States
\item \Idef{org141}Westf\"{a}lische Wilhelms-Universit\"{a}t M\"{u}nster, Institut f\"{u}r Kernphysik, M\"{u}nster, Germany
\item \Idef{org142}Wigner Research Centre for Physics, Hungarian Academy of Sciences, Budapest, Hungary
\item \Idef{org143}Yale University, New Haven, Connecticut, United States
\item \Idef{org144}Yonsei University, Seoul, Republic of Korea
\end{Authlist}
\endgroup
\end{document}